\documentclass[12pt,preprint]{aastex}
\usepackage{amssymb}
\usepackage{epsfig}
\usepackage{lscape}
\usepackage{longtable} 
\usepackage{url}
\usepackage{array}

\newcommand{\etal}{et~al.~}

\begin{document}


\title{Optical Spectroscopy of H$\alpha$ Filaments in Cool Core Clusters: Kinematics, Reddening and Sources of Ionization}

\author{Michael McDonald\altaffilmark{1,4}, Sylvain Veilleux\altaffilmark{2,5} and David S. N. Rupke\altaffilmark{3,6}}

\altaffiltext{1} {Kavli Institute for Astrophysics and Space Research, MIT, Cambridge, MA 02139, USA}
\altaffiltext{2}{Department of Astronomy, University of Maryland, College
  Park, MD 20742, USA} 
\altaffiltext{3}{Department of Physics, Rhodes College, Memphis, TN 38112, USA}  
\altaffiltext{4}{Email: mcdonald@space.mit.edu}
\altaffiltext{5}{Email: veilleux@astro.umd.edu}
\altaffiltext{6}{Email: rupked@rhodes.edu}


\begin{abstract}

We have obtained deep, high spatial and spectral resolution, long-slit spectra of the H$\alpha$ nebulae in the cool cores of 9 galaxy clusters. This sample provides a wealth of information on the ionization state, kinematics, and reddening of the warm gas in the cool cores of galaxy clusters. We find evidence for only small amounts of reddening in the extended, line-emitting filaments, with the majority of filaments having $E(B-V)<0.2$. We find, in agreement with previous works, that the optical emission in cool core clusters has elevated low-ionization line ratios. The combination of [\ion{O}{3}]/H$\beta$, [\ion{N}{2}]/H$\alpha$, [\ion{S}{2}]/H$\alpha$, and [\ion{O}{1}]/H$\alpha$ allow us to rule out collisional ionization by cosmic rays, thermal conduction, and photoionization by ICM X-rays and AGN as strong contributors to the ionization in the bulk of the optical line-emitting gas in both the nuclei and filaments. The data are adequately described by a composite model of slow shocks and star formation. This model is further supported by an observed correlation between the linewidths and low ionization line ratios which becomes stronger in systems with more modest star formation activity based on far ultraviolet observations.  We find that the more extended, narrow filaments tend to have shallower velocity gradients and narrower linewidths than the compact filamentary complexes. We confirm that the widths of the emission lines decrease with radius, from FWHM $\sim 600$ km s$^{-1}$ in the nuclei to FWHM $\sim 100$ km s$^{-1}$ in the most extended filaments. The variation of linewidth with radius is vastly different than what is measured from stellar absorption lines in a typical giant elliptical galaxy, suggesting that the velocity width of the warm gas may in fact be linked to ICM turbulence and, thus, may provide a glimpse into the amount of turbulence in cool cores. In the central regions ($r<10$ kpc) of several systems the warm gas shows kinematic signatures consistent with rotation, consistent with earlier work. We find that the kinematics of the most extended filaments in this sample are broadly consistent with both infall and outflow, and recommend further studies linking the warm gas kinematics to both radio and X-ray maps in order to further understand the observed kinematics. 

\end{abstract}


%================================================================%
%============== INTRODUCTION ====================================%
%================================================================%

\section{Introduction}
The dense core of a galaxy cluster represents a unique environment, where the hot intracluster medium (ICM) is cooling most rapidly, feedback from the central active galactic nucleus (AGN) is most effective, and the brightest cluster galaxy (BCG) dominates the mass. This represents one of the few places in the Universe where large-scale cooling and feedback processes can be readily observed. Unlike isolated massive galaxies, where some of the energy injected into the interstellar medium (ISM) from the AGN often escapes into the low-density intergalactic medium (IGM), the denser ICM in cluster cores retains an imprint of this feedback in the form of bubbles \citep[e.g., ][]{churazov01, reynolds05, revaz08} or ripples \citep[e.g., Perseus A; ][]{fabian03}. Similarly, the accretion of hot gas from the IGM onto massive galaxies is challenging to observe due to the very low densities, while such phenomena have been studied in depth for decades in galaxy clusters. The so-called ``cooling flows'' in galaxy clusters, which were once thought to be massive flows of cool gas on the order of 100--1000~M$_{\odot}$~yr$^{-1}$ \citep[see review by ][]{fabian94}, are now understood to be considerably less massive, depositing on the order of 1--10 M$_{\odot}$~yr$^{-1}$ of cool gas onto the BCG \citep[e.g., ][]{voigt04, bregman06}. It is assumed that some form of feedback (e.g., AGN, gas sloshing, conduction) prevents the dense ICM from catastrophically cooling, allowing only a trickle of cool gas to accrete onto the BCG. 

There is a considerable amount of support for the ``reduced cooling flow'' model \citep{voigt04} from observations at a variety of wavelengths. Cool core clusters  (galaxy clusters for which the central ICM cooling time is much less than the age of the Universe) are well known to have an abundance of multi-phase gas in the central $\sim 100$~kpc which has been observed in O VI \citep[e.g., ][]{bregman01, oegerle01a, bregman06}, H$\alpha$ \citep[e.g., ][]{hu85, johnstone87, heckman89, crawford99, jaffe05, edwards07,hatch07, mcdonald10, mcdonald11a}, and H$_2$ \citep[e.g., ][]{edge01, salome03, jaffe05, lim08, donahue11}. Additionally, evidence for star formation has been observed in the UV \citep[e.g., ][]{mcnamara89, rafferty06, mcdonald09, hicks10, mcdonald11b}, optical \citep[e.g., ][]{allen95, cardiel95, crawford99, edwards07, bildfell08}, and mid IR \citep[e.g., ][hereafter MIR]{hansen00, egami06, odea08, quillen08}. In \citet{mcdonald11b}, we showed that the star formation rates inferred from far-UV and mid-IR data suggest an efficiency of $\sim$ 15\% in converting the cooling ICM into stars, consistent with the global baryon fraction in stars. This star formation efficiency estimate, based on far-UV, H$\alpha$, and mid-IR data, is consistent with earlier work based on mid-IR star formation estimates by \cite{egami06} and \cite{odea08}. This overwhelming evidence for cooling byproducts suggests that cooling flows are occurring, but on a reduced scale than was initially predicted.

In \citet{mcdonald10,mcdonald11a} we presented a sample of 33 galaxy groups and clusters with deep, high-spatial-resolution H$\alpha$ imaging from the Maryland-Magellan Tunable Filter \citep[MMTF\footnote{http://www.astro.umd.edu/~veilleux/mmtf/};][]{veilleux10}. We reported on several new systems with extended, filamentary H$\alpha$ emission, and produced higher-quality H$\alpha$ maps for some previously known systems. By combining deep, archival Chandra X-ray Observatory (CXO) data with these new H$\alpha$ images, we showed that the extended warm gas (H$\alpha$) is spatially coincident with the asymmetric, rapidly-cooling ICM for an ensemble of optically-emitting nebulae and that the H$\alpha$ emission is always confined within the ICM cooling radius. This, taken together with a number of previously known correlations between the warm and hot phases \citep[see review by][]{fabian94}, showed that the H$\alpha$ emission is intimately linked to the X-ray cooling flow in both quantity and morphology. 
However, while we were able to determine the origin of this cool gas (cooling out of the hot ICM), we were unable to constrain the dominant ionization processes which produce the high H$\alpha$ fluxes. In an attempt to shed light on this remaining mystery, we have acquired deep, high-resolution spectra of multiple filaments and BCG nuclei in a sample of cool core clusters drawn from our previous works \citep{mcdonald10,mcdonald11a,mcdonald11b}. We present these data in \S2, describing their collection, reduction, and analysis. In \S3 we present the results of our analysis of the spectra, providing new information on the kinematics, ionization, and reddening of the warm, ionized gas filaments in cluster cores. In \S4 we compare these results to various models in an attempt to ascertain the source of ionization in these filaments. Finally, we summarize our results in \S5 and discuss future work which may shed further light on this topic.

Throughout this paper, we assume the following cosmological values: H$_0$ = 73 km s$^{-1}$ Mpc$^{-1}$, $\Omega_{matter}$ = 0.27, $\Omega_{vacuum}$ = 0.73.

%\vspace{5in}

%================================================================%
%============== DATA ============================================%
%================================================================%

\section{Data Collection and Analysis}

\begin{table*}[htb]
\begin{center}
\begin{tabular}{c c c c c c}
\hline\hline
Name & z & E(B-V) & Source & Slit PAs & H$\alpha$ Ref.\\
(1) & (2) & (3) & (4) & (5) & (6)\\
\hline
 Abell~0478 & 0.0881 & 0.517 & Keck -- LRIS & 356.$^{\circ}$4, 6.$^{\circ}$4& M10\\
 Abell~0496 & 0.0329 & 0.132 & Keck -- LRIS & 77.$^{\circ}$4, 40.$^{\circ}$4& M10\\
 Abell~0780 & 0.0539 & 0.042 & Magellan -- IMACS & 110.$^{\circ}$0, 72.$^{\circ}$0 & M10\\
 Abell~1644 & 0.0475 & 0.069 & Magellan -- IMACS & 159.$^{\circ}$0, 79.$^{\circ}$0 & M10\\
 Abell~1795 & 0.0625 & 0.013 & Magellan -- IMACS & 350.$^{\circ}$4, 344.$^{\circ}$4 & M10\\
 Abell~2052 & 0.0345 & 0.037 & Magellan -- IMACS & 258.$^{\circ}$1, 108.$^{\circ}$0 & M10\\
 Abell~2597 & 0.0830 & 0.030 &  Magellan -- IMACS & 14.$^{\circ}$0 & M11b\\
 NGC~4325 & 0.0257 & 0.023 & Magellan -- IMACS & 33.$^{\circ}$,5 163.$^{\circ}$2 & M11a\\
 Sersic~159-03  & 0.0580 & 0.011 & Magellan -- IMACS & 86.$^{\circ}$0, 161.$^{\circ}$9 & M10,11a\\
\hline

\end{tabular}
\end{center}
\caption{Properties of our sample of 9 galaxy clusters with long-slit spectroscopy. The columns are: (1) Cluster name; (2) BCG redshift; (3) Galactic reddening toward cluster center from \citet{schlegel98}; (4) Source of long-slit data; (5) Position angles of slits in units of degrees east of north; (6) Reference for MMTF H$\alpha$ data.}
\label{sample}
\end{table*}

\begin{figure}[p]
\centering
\includegraphics[width=0.8\textwidth]{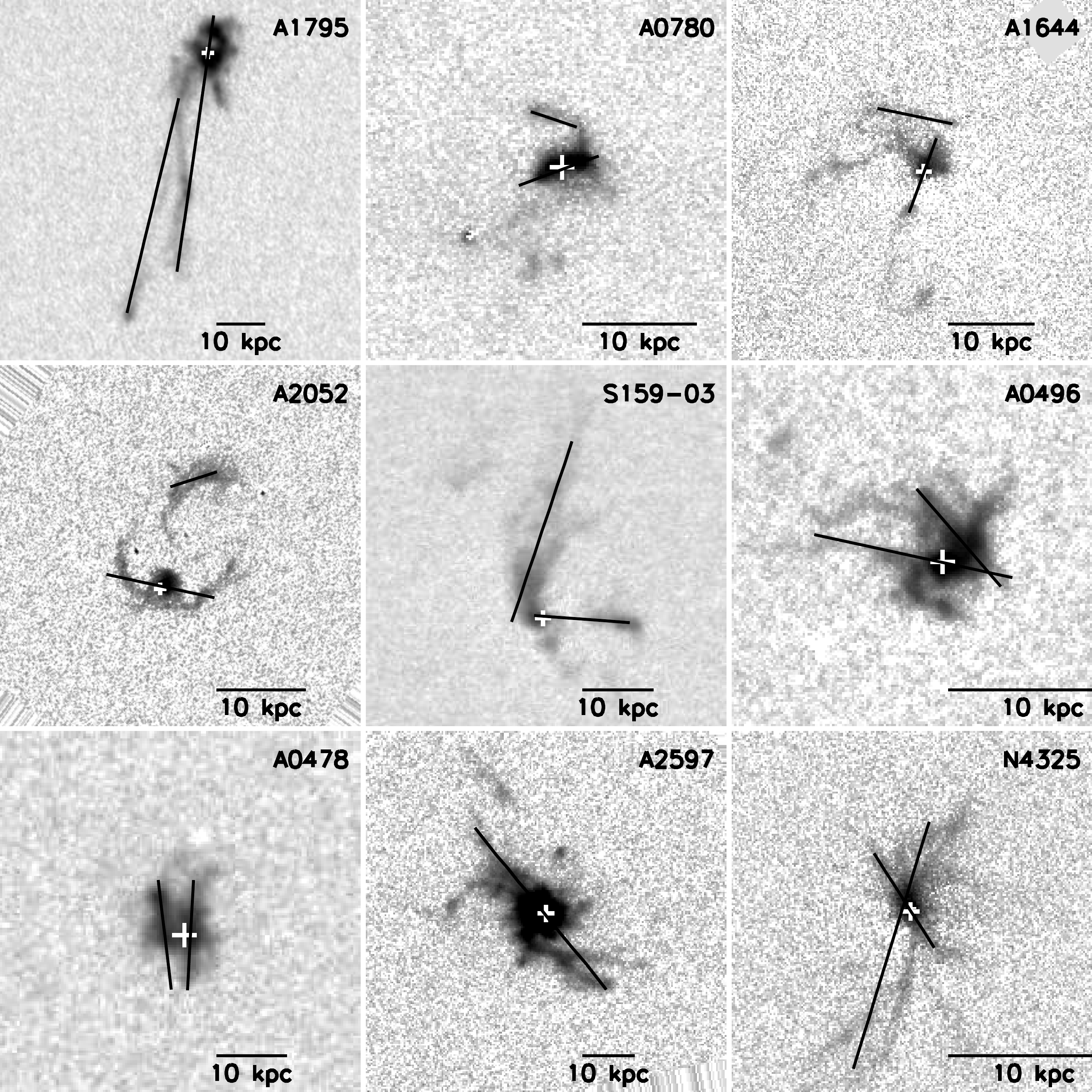}
\caption{Slit positions for all 9 cluster cores in our sample overlaid on H$\alpha$ images from \cite{mcdonald10,mcdonald11a,mcdonald11b}. The length of the slits have been artificially truncated for this plot to cover only optically-emitting regions. In all cases we have attempted to maximize the overlap between these slits and the H$\alpha$ maps while forcing one slit to pass through the BCG nucleus. In the case of Abell~2597 we were only able to obtain one pointing due to time constraints. In NGC~4325 one of the slits is slightly offset from the ideal position due to a pointing error.}
\label{3x3_slits}
\end{figure}

In order to investigate the various properties of the optically-emitting filaments in cool cores, we performed long-slit spectroscopy on a sample of 9 BCGs.  This sample was chosen to include the most extended and luminous systems at H$\alpha$ from our larger sample of 36 clusters with deep H$\alpha$ imaging from the Maryland-Magellan Tunable Filter \citep{veilleux10} as described in \citet{mcdonald10,mcdonald11a,mcdonald11b} (Table \ref{sample}). Long-slit spectra were obtained over the span of 2 years at the Magellan and Keck telescopes, using the Inamori-Magellan Areal Camera \& Spectrograph \citep[IMACS;][]{imacs} and Low Resolution Imaging Spectrometer \citep[LRIS;][]{lris}, respectively. The setup we used allowed for a broad wavelength coverage, allowing full spectral coverage from $\sim$5000--8000\AA\ with IMACS and $\sim$3500--7500\AA\ with LRIS, and relatively high dispersions (IMACS: 0.385 \AA/pix @ 7000\AA, LRIS: 0.537 \AA/pix at @ 7000\AA), allowing for careful linewidth measurements down to the intrinsic instrumental linewidth (FWHM $\sim$ 75 and 145 km s$^{-1}$ for IMACS and LRIS @ 7000\AA, respectively). For each cluster, two slits were aligned along the filaments in an attempt to maximize the coverage of the optically-emitting gas. Additionally, one of the two slits was always forced to pass through the BCG nucleus. In the case of Abell~2597 only a single slit position was acquired due to time constraints, and in NGC~4325 one of the two slits is slightly offset from the optimal position.  The chosen slit positions are shown in Fig. \ref{3x3_slits}. 

% Figure with ~2-D Spectra

In total, 17 optical spectra of 9 cluster cores were obtained. These spectra were reduced using standard \emph{IRAF} (\url{http://iraf.net}) packages for the reduction of long-slit spectra. This procedure involved removing the zero-point bias from and flat-fielding each exposure (CCDPROC), masking cosmic rays (LA Cosmic; \url{http://www.astro.yale.edu/dokkum/lacosmic/}), combining similar exposures (IMCOMBINE), removing spatial and spectral distortions (TRANSFORM), removing sky lines (XVista; \url{http://ganymede.nmsu.edu/holtz/xvista/}), and calibrating the spectra in both wavelength (IDENTIFY, FITCOORDS) and intensity (STANDARD, SENSFUNC, CALIBRATE). The wavelength calibration was derived from spectra of helium, neon, and argon arc lamps, while the photometric calibration was based on the spectrophotometric flux standards EG274, LTT3218, and G191-B2B.  Spectra were visually inspected for contamination of the emission lines from the $O_2$ atmospheric absorption features around 6870\AA\ and 7620\AA.

From each spectrum, the redshift and velocity dispersion of the stars at the center of the BCG were measured using the Na D lines at 5890\AA\ and 5896\AA\ and the \ion{Mg}{1} line at 5178\AA\ wherever available. Next, the H$\alpha$ and [\ion{N}{2}] lines, which do not require deconvolution due to the high spectral resolution, were simultaneously fit using a combination of three Gaussians. It was assumed that all three lines had the same velocity dispersion and redshift, which yields 4 free parameters: $v, \sigma_v, F_{H\alpha}, F_{[N~II]}$. In the few cases where multiple velocity components were visible, we fit two Gaussians to each line.  The continuum was fit using a three-parameter model, which consisted of a linear component (2 parameters) and an H$\alpha$ stellar absorption feature with the width and redshift fixed at the value measured from the Na D and \ion{Mg}{1} lines. All spectra were visually inspected for contamination of the emission lines from the $O_2$ atmospheric features at 6870\AA\ and 7620\AA. Of our 9 targets, only Abell~1644 is redshifted such that the H$\alpha$+[\ion{N}{2}] lines are affected by $O_2$ absorption. For this system, we clipped the spectrum from 6870--6876\AA, which removed the strongly-absorbed part of the spectrum while leaving the majority of the emission lines intact. We note that, while the H$\alpha$ line fluxes in this system are likely still contaminated and should be taken as lower limits, they do not appear as outliers in any of the observed trends and have H$\beta$ fluxes and H$\alpha$ FWHM consistent with the low observed flux, as we will show.

Following the fits to the H$\alpha$+[\ion{N}{2}] lines, the H$\beta$, [\ion{O}{3}], [\ion{O}{1}], and [\ion{S}{2}] lines were individually isolated and fit with Gaussians. For each of these fits the redshift and velocity dispersion were fixed to the values for H$\alpha$, while the line intensity and continuum zeropoint and slope were allowed to vary. Additionally, an H$\beta$ stellar absorption feature was included with a variable amount of absorption. Finally, a visual inspection of all 5 fitting regions was performed for each spectrum and spurious fits were re-run with more specific input parameters, allowing for a more reliable fit. All measured line intensities were further corrected for Galactic extinction following \citet{cardelli89} using reddening estimates from \citet{schlegel98}.

%================================================================%
%============== RESULTS =========================================%
%================================================================%
\section{Results}

These data, described in full in Appendix A, provide a wealth of information about the warm gas kinematics, ionization state, and amount of reddening in the cool cores of galaxy clusters. In this section, we first address the results of our spectroscopic analysis in the context of these three major topics, and then follow with a discussion of the implications that these results have on our current understanding of the generation of these optical filaments. 

\subsection{Optical Line Ratios}

% Figure with ~2-D Spectra
\begin{figure*}[p]
\centering
\includegraphics[width=0.99\textwidth]{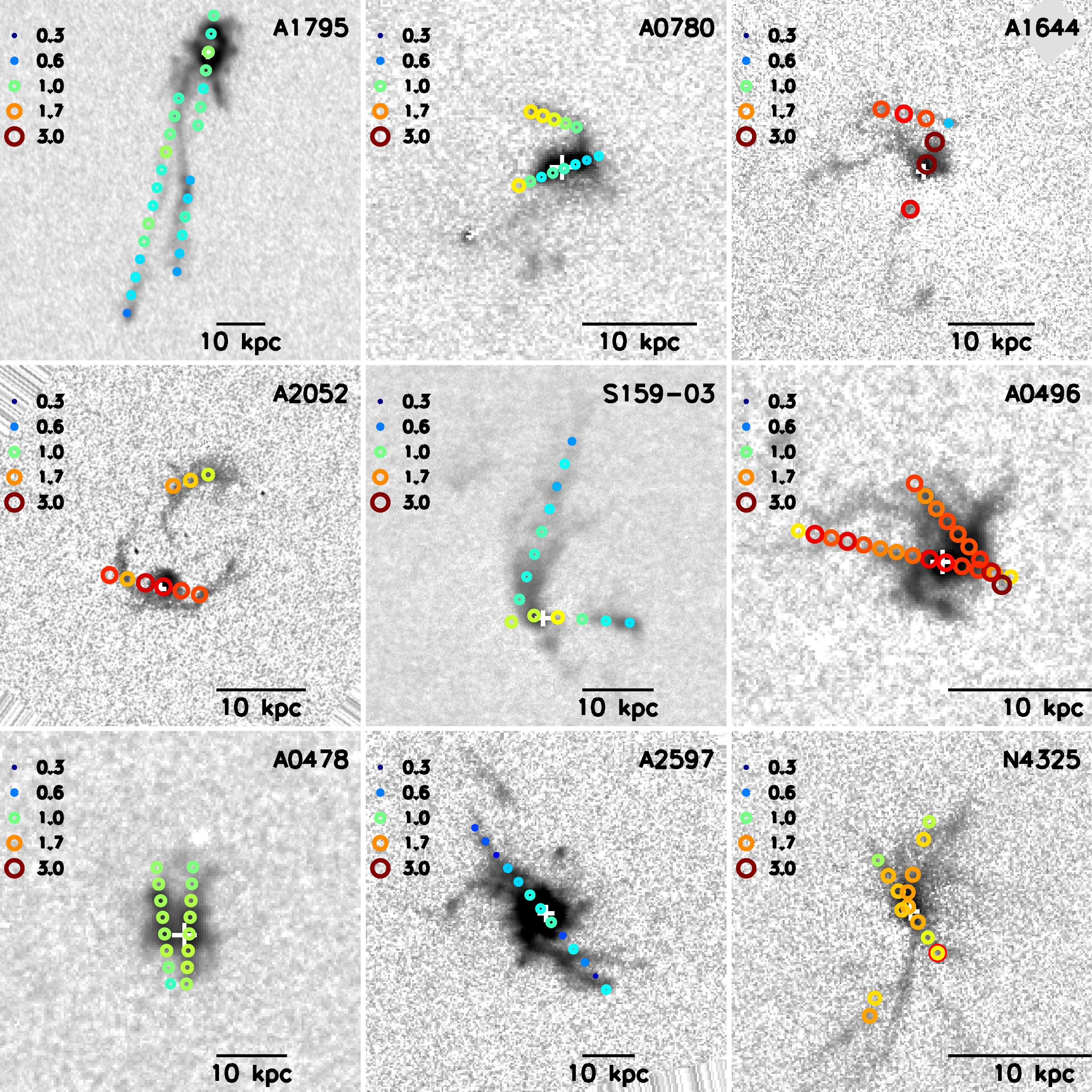}
\caption{Spatial distribution of the [\ion{N}{2}] $\lambda$6583/H$\alpha$ intensity ratio. The combination of narrow-band H$\alpha$ imaging with long-slit spectroscopy allows us to create pseudo-2D spectra. In all panels, both the point size and color correspond to the magnitude of the line ratio, as described by the legend in the upper left. The cluster or central galaxy name is shown in the upper right, while the physical scale of the image is shown via a 10kpc scale bar in the lower right. The white cross represents the center of the optical (stellar continuum) emission. }
\label{3x3_n2h}
\end{figure*}

\begin{figure*}[p]
\begin{tabular}{c c}
\includegraphics[width=0.48\textwidth]{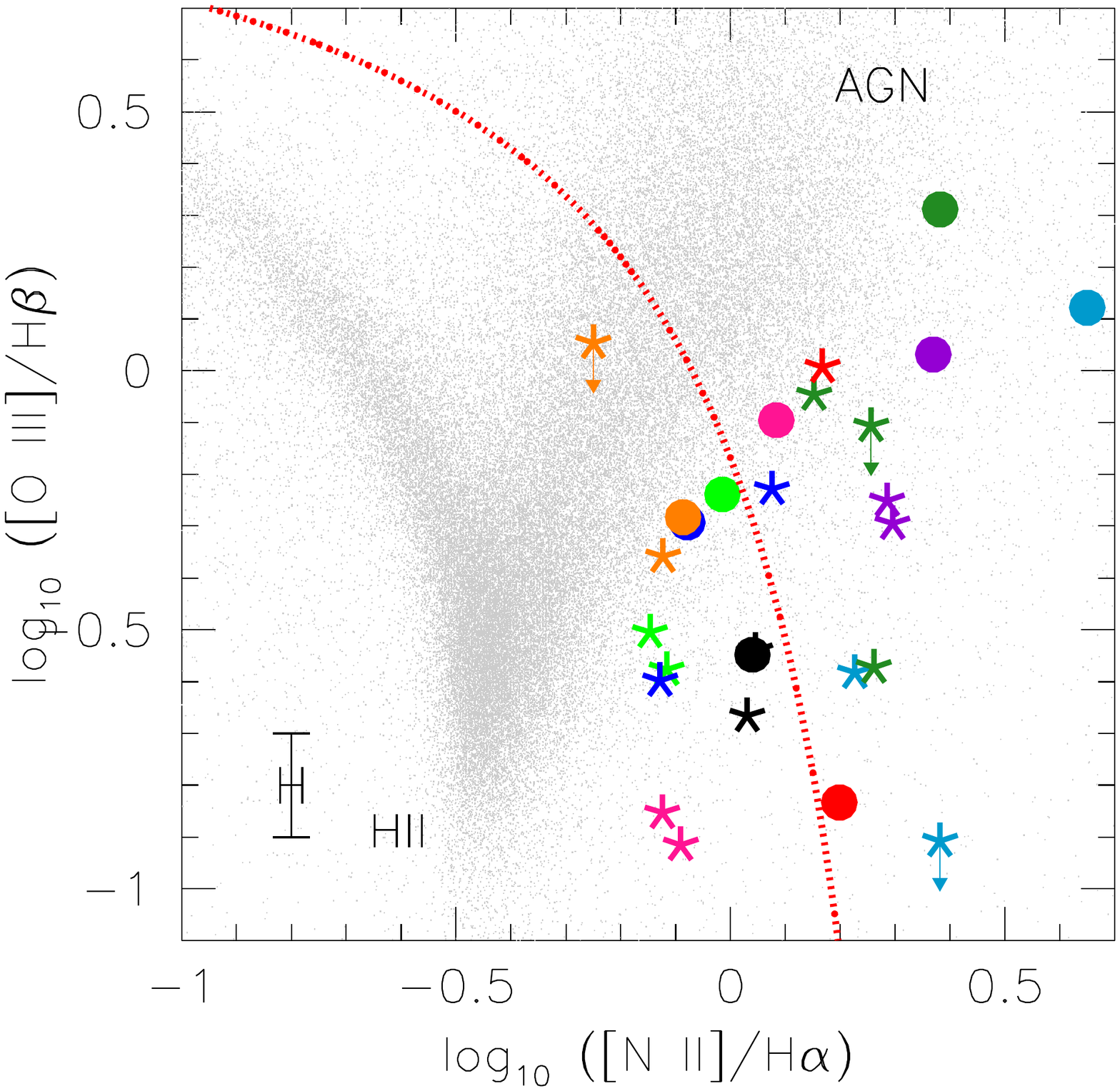} & 
\includegraphics[width=0.48\textwidth]{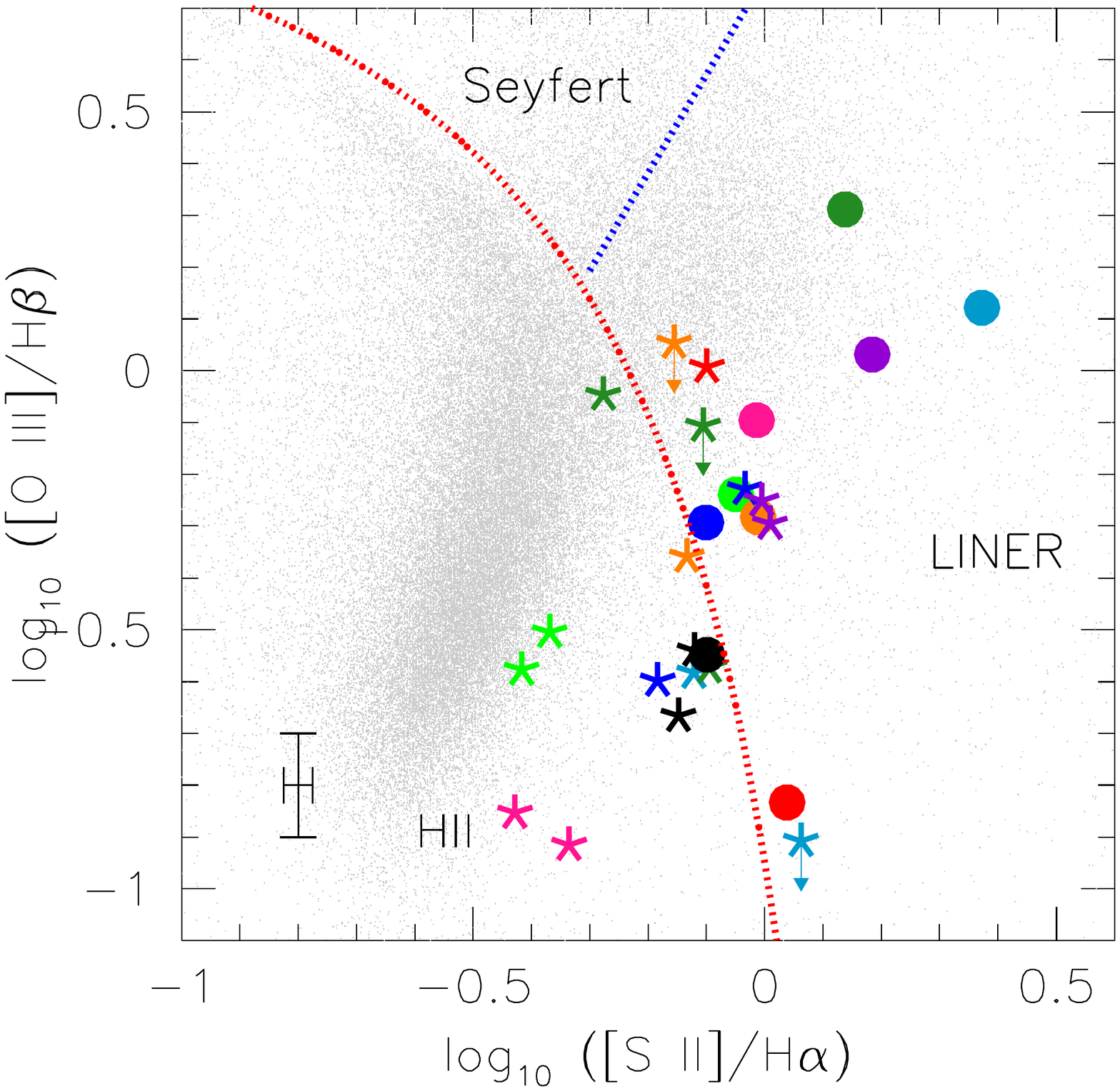}\\
\includegraphics[width=0.48\textwidth]{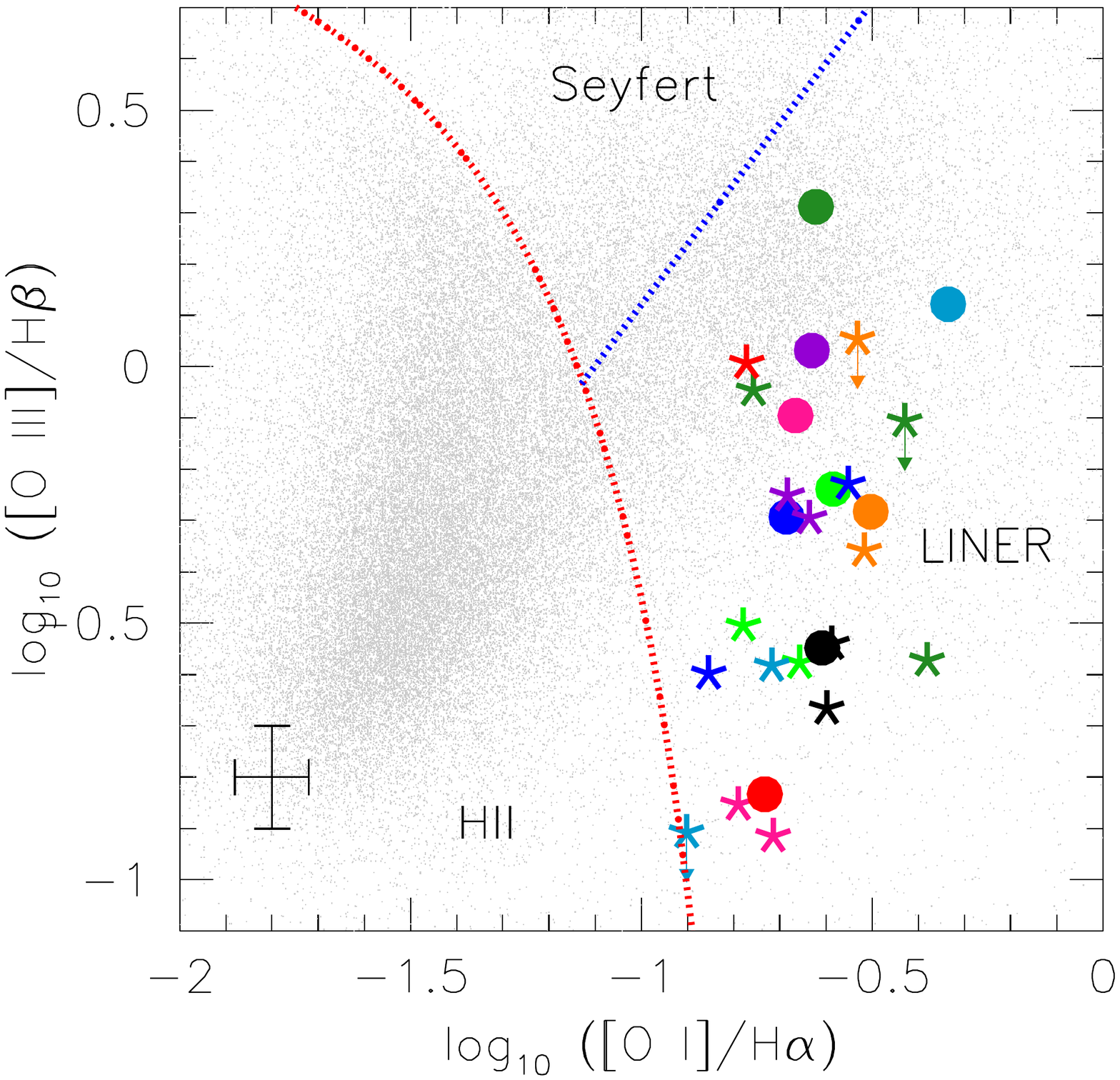} & 
\includegraphics[width=0.48\textwidth, trim=0mm 0mm 20mm 0mm]{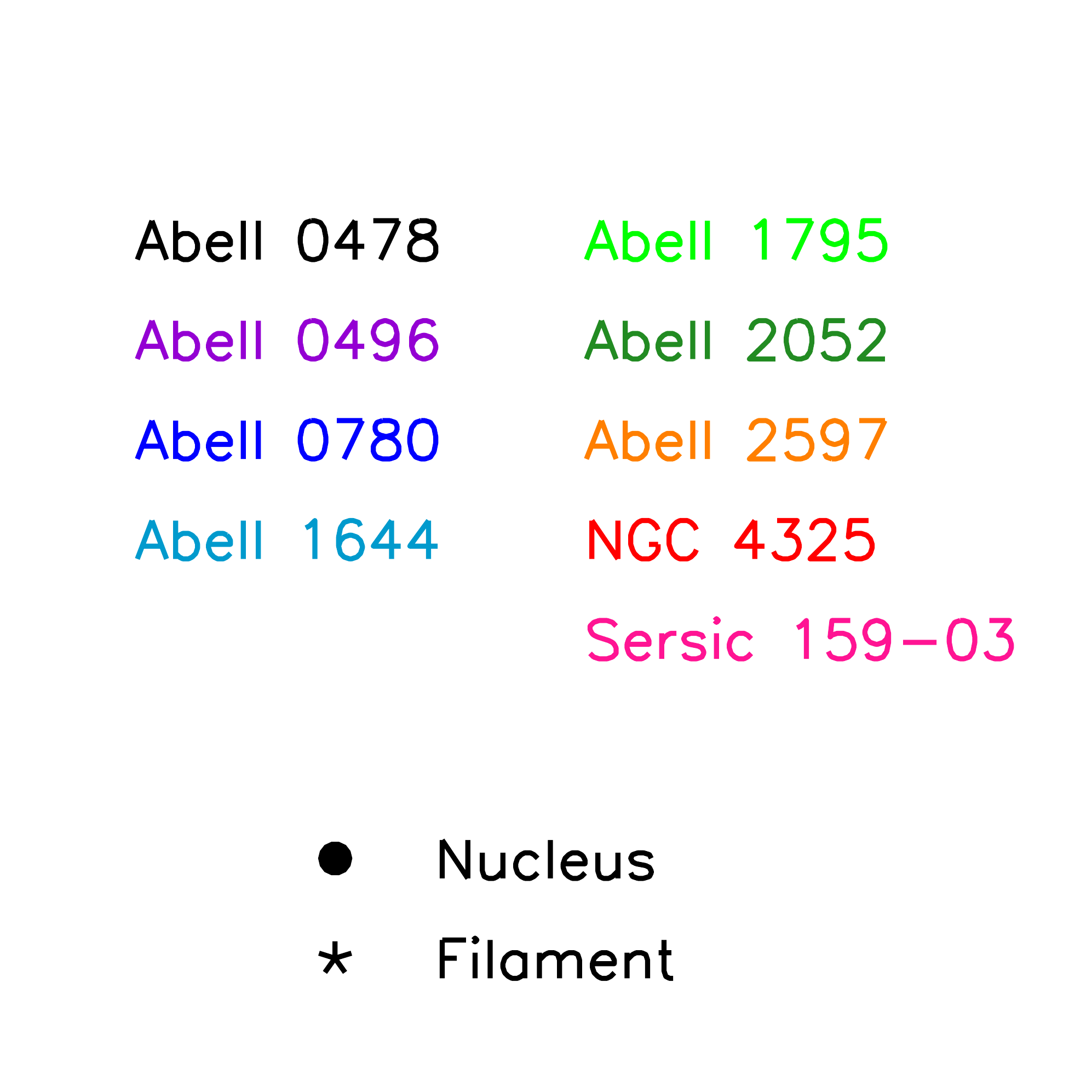} \\
\end{tabular}
\caption{Key diagnostic diagrams involving the reddening-insensitive line ratios [\ion{N}{2}] $\lambda$6583/H$\alpha$, [\ion{O}{1}] $\lambda$6300/H$\alpha$, [\ion{S}{2}] $\lambda\lambda$6716, 6731/H$\alpha$, and [\ion{O}{3}] $\lambda$5007/H$\beta$ for filaments (stars) and nuclei (circles) in each of our 9 systems. The grey points are galaxies from the Sloan Digital Sky Survey \citep{kewley06} and show a clear separation into \ion{H}{2} regions and AGN. The red line represents the extreme starburst limit and separates HII regions from AGN, while the blue line separates AGN into Seyfert and LINER classes \citep{kewley06}. Typical errorbars are shown in the bottom left corner of each panel. We note that, in general, the data for optical emission in cool core clusters tend towards elevated low-ionization lines.}
\label{lineratios}
\end{figure*}

The use of deep, high spatial resolution, narrow-band H$\alpha$ imaging to dictate the orientation of our long-slit spectroscopy allows us to characterize the properties of the optical emission in two spatial dimensions without the field-of-view limitations of 3-D spectroscopy.
The broad wavelength coverage of these spectra allows the comparison of various optical emission lines, specifically the key diagnostic line ratios [\ion{N}{2}] $\lambda$6583/H$\alpha$, [\ion{O}{1}] $\lambda$6300/H$\alpha$, [\ion{S}{2}] $\lambda\lambda$6716, 6731/H$\alpha$, and [\ion{O}{3}] $\lambda$5007/H$\beta$. This combination of line ratios is traditionally used to identify the source of ionization in emission-line regions \citep{baldwin81,veilleux87,kewley06}. We show in Figure \ref{3x3_n2h} the first of these ratios, [\ion{N}{2}]/H$\alpha$, which is often used to differentiate between star-forming regions and AGN, overlaid on H$\alpha$ maps of individual cluster cores. Only 2/9 clusters (Abell~1644, Abell~2052) have [\ion{N}{2}]/H$\alpha$ strongly peaked on the nucleus, where we would expect AGN, if present, to dominate the ionization. Roughly 2/3 of systems have low ($\lesssim 1.5$) [\ion{N}{2}]/H$\alpha$ everywhere, while the remaining 1/3 of systems have high ($\gtrsim 1.5$) [\ion{N}{2}]/H$\alpha$ everywhere.  Interestingly, this is the same fraction of cool cores that appear to be star forming \citep{mcdonald11b}. For systems with both far-UV imaging from \cite{mcdonald09} and \cite{mcdonald11b} and optical spectroscopy, we find a one-to-one correspondence between star-forming (Abell~1795,  Abell~2597) and shock-heated (Abell~1644, Abell~2052) systems based on [\ion{N}{2}]/H$\alpha$ and far-UV/H$\alpha$ flux ratios. The systems with the longest filaments (Abell~1795, Abell~2597, Sersic~159-03) have the lowest overall [\ion{N}{2}]/H$\alpha$ ratios, with the value of the ratio decreasing slightly at larger radius. However, these [\ion{N}{2}]/H$\alpha$ ratios are significantly elevated above what one would expect for ongoing star formation ($\sim 0.3$ for solar metallicity), but are roughly consistent with the expectations for starburst systems \citep[$\sim 1.0$; e.g., ][]{kewley06}. Several prior studies have also found elevated low-ionization line ratios in the optical emission in cool core clusters \citep[e.g., ][]{voit97,crawford99}. In order to understand these line ratios, we appeal to the full suite of line ratios first used by \citet{veilleux87}.

In Figure \ref{lineratios}, we show the [\ion{O}{3}]/H$\beta$ line ratio as a function of [\ion{N}{2}]/H$\alpha$, [\ion{O}{1}]/H$\alpha$, and [\ion{S}{2}]/H$\alpha$. In these plots we show data from the Sloan Digital Sky Survey \citep[SDSS; ][]{sdss} for emission-line nuclei, showing the separate regions occupied by star-forming galaxies, low-ionization nuclear emission-line regions (LINERs), and Seyfert galaxies. We find a systematic offset between the emission-line ratios in the warm filaments and the locus of points from the SDSS, towards lower [\ion{O}{3}]/H$\beta$ and/or higher [\ion{N}{2}]/H$\alpha$, [\ion{S}{2}]/H$\alpha$, and [\ion{O}{1}]/H$\alpha$. We note that the grey points represent \emph{nuclei}, while much of our data includes a stronger contribution from diffuse, ionized regions. The fact that excess blue light  \citep[e.g., ][]{crawford99,edwards07,bildfell08}, UV emission \citep[e.g., ][]{rafferty06,hicks10,mcdonald11b}, mid-far IR emission \citep[e.g., ][]{odea08}, and molecular gas \citep[e.g., ][]{edge01,salome03} are observed is strong evidence for star formation, but  Figure \ref{lineratios} seems to indicate that the situation is more complex. The [\ion{N}{2}]/H$\alpha$ and [\ion{S}{2}]/H$\alpha$ ratios in some filaments are consistent with the extreme starburst limit \citep{kewley06}, however the [\ion{O}{1}]/H$\alpha$ ratio is far too high to be pure star formation. Thus, it appears that pure star formation, whether ongoing or burst-like, is unable to explain all of the observed optical line ratios in the cool cores of galaxy clusters. In \S4.1, we consider different combinations of ionization mechanisms which can explain the elevated low-ionization ratios (Figure \ref{lineratios}) while still being consistent with the observations at other wavelengths that seem to strongly suggest the presence of star formation.

The nuclei in our sample (colored circles in Figure \ref{lineratios}) have systematically low [\ion{O}{3}]/H$\beta$, consistent with LINERs. This classification is consistent with the relative weak luminosity of nuclei in cool core BCGs in both the optical and X-ray and with the narrow emission lines (Figure \ref{sigr}). However, only three nuclei (Abell~0496, Abell~1644, and Abell~2052) are clearly offset from the filaments in Figure \ref{lineratios}. The remaining six nuclei are likely undergoing the same ionization processes as the filaments and, thus, may not be AGN, or may be composite objects \citep{kewley06}, despite their identification as LINERs in Figure \ref{lineratios}.

\begin{figure}[htb]
\centering
\includegraphics[width=0.6\textwidth]{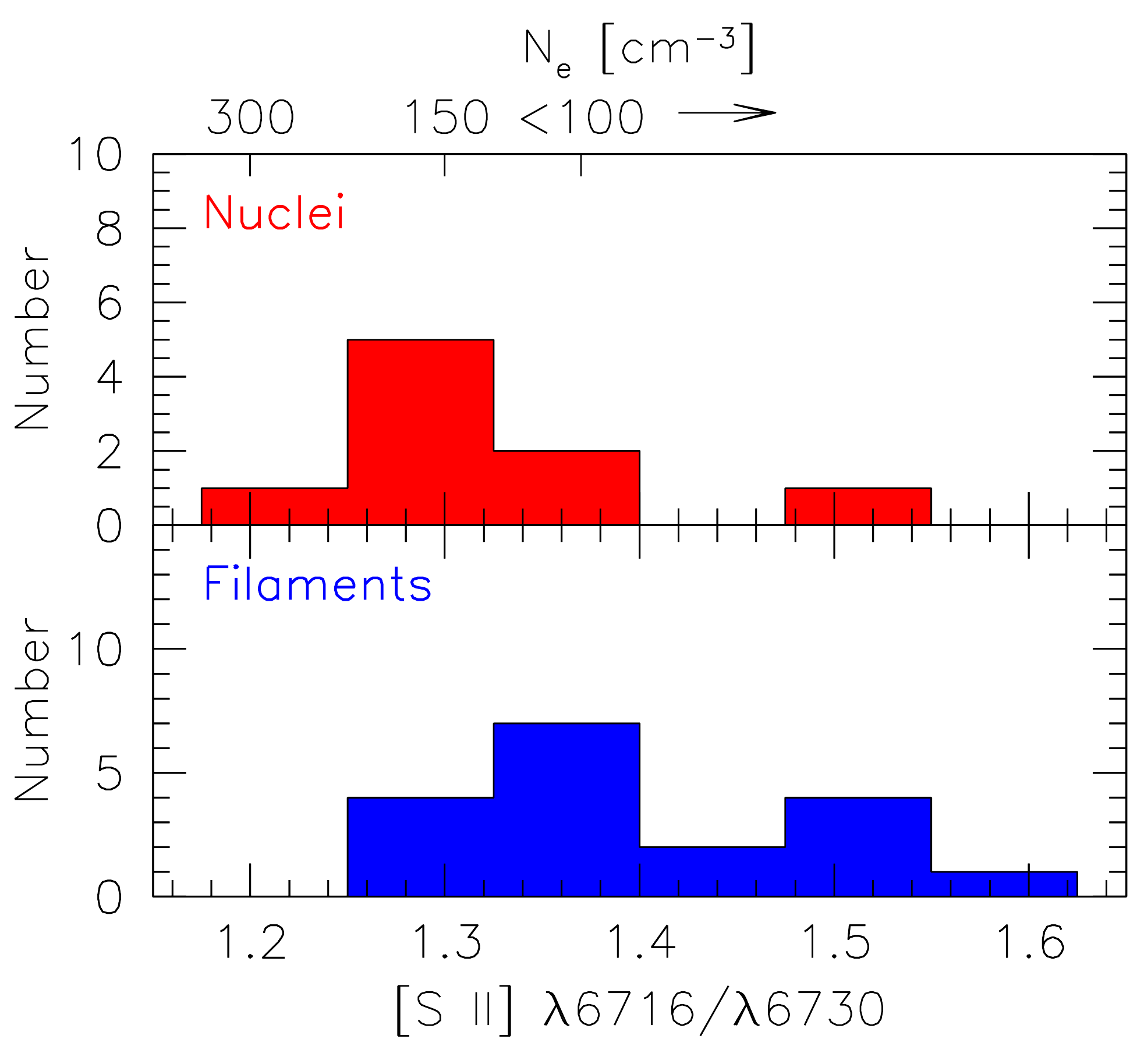}
\caption{Distribution of the [\ion{S}{2}] $\lambda$6716/$\lambda$6730 line ratio in the nuclei (upper) and filaments (lower) of 9 cool core clusters. The upper axis shows the inferred electron density of the optical line-emitting gas. At densities below 100 cm$^{-3}$, the [\ion{S}{2}] $\lambda$6716/$\lambda$6730 line ratio becomes ineffective and can only provide an upper limit. }
\label{sii_hist}
\end{figure}

The high spectral resolution of these data allows us to separate the [\ion{S}{2}] doublet and determine the [\ion{S}{2}] $\lambda$6716/$\lambda$6730 intensity ratio, which can then be used to probe the density of the line-emitting gas, as described in \cite{osterbrock89}. In Figure \ref{sii_hist}, the distribution of the [\ion{S}{2}] $\lambda$6716/$\lambda$6730 ratio is shown for both filaments and nuclei in our sample. The line-emitting gas in nuclei has typical densities between 100--300 cm$^{-3}$, consistent with earlier work by \cite{heckman89}. The more extended gas in the filaments appears to have overall lower densities, although we are unable to determine the average density due to the inability of the [\ion{S}{2}] $\lambda$6716/$\lambda$6730 ratio to probe densities below $\sim$100 cm$^{-3}$. In general, the filaments tend to have $N_e<200$ cm$^{-3}$, potentially reaching much lower than $\sim$100 cm$^{-3}$ at large radius.

\begin{figure}[htb]
\centering
\includegraphics[width=0.5\textwidth]{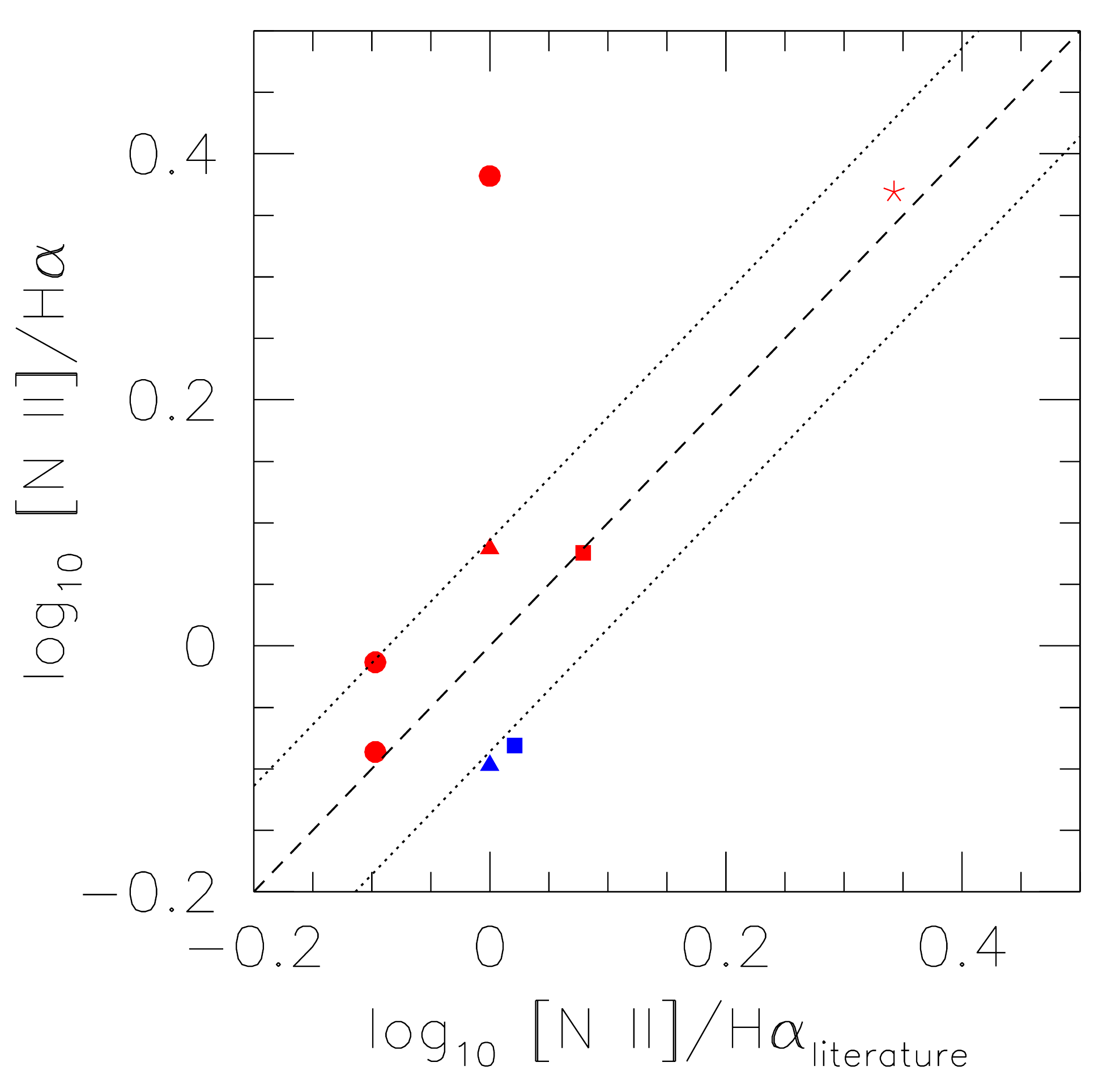}
\caption{Comparison of [\ion{N}{2}]/H$\alpha$ line ratios for this work to those taken from the literature. Blue and red points represent extended and nuclear emission, respectively. Point types correspond to different references: \cite{heckman89} -- filled circles, \cite{melnick97} -- filled squares, \cite{jaffe05} -- filled triangles, \cite{hatch07} -- stars. The dashed line represents equality, while the dotted lines represent a scatter of $\sim$20\%. The significant outlier is Abell~2052, for which \cite{heckman89} had to decompose the blended [\ion{N}{2}] and H$\alpha$ lines.}
\label{compare_n2h}
\end{figure}

\begin{figure}[htb]
\centering
\includegraphics[width=0.8\textwidth]{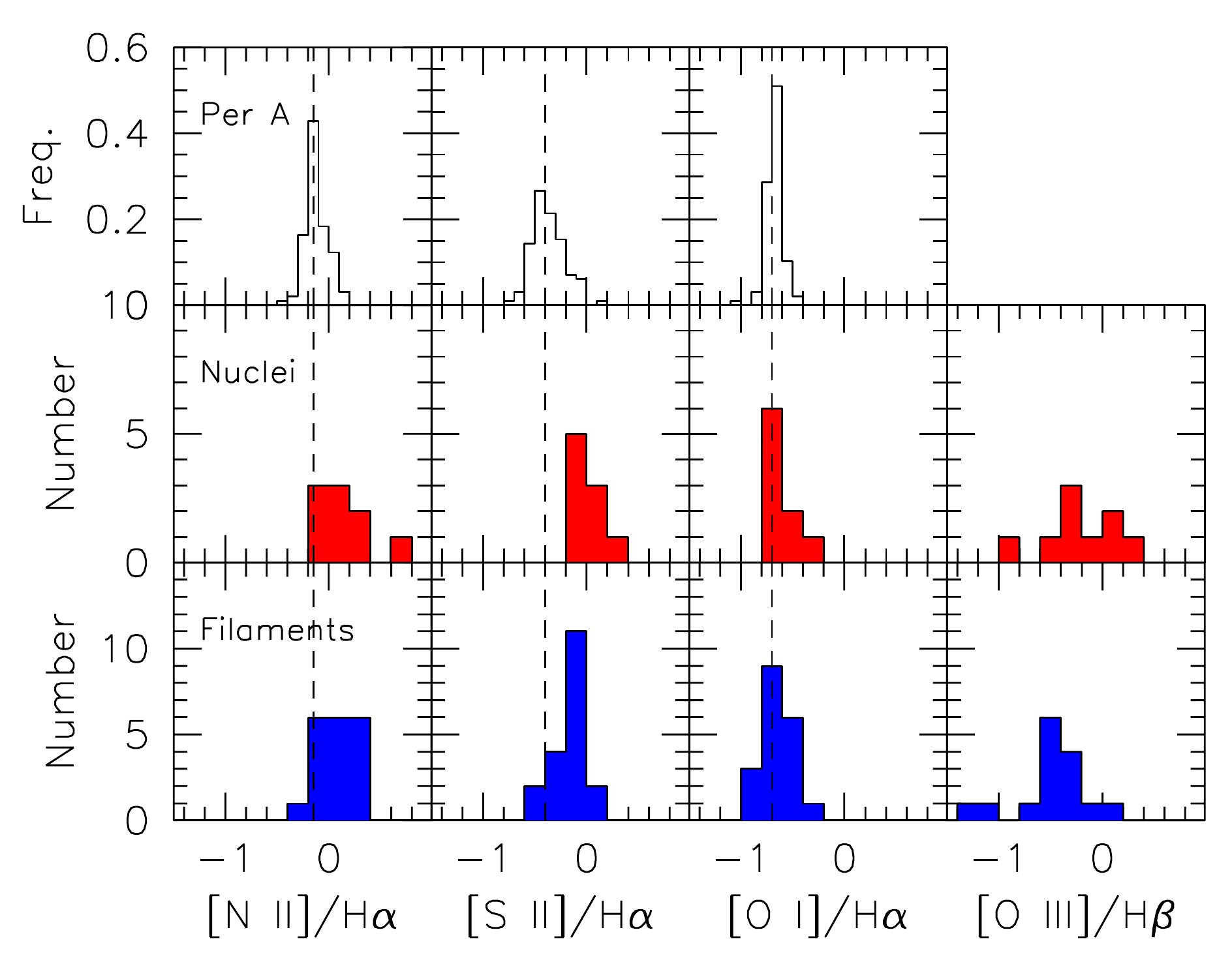}
\caption{Distribution of reddening-insensitive optical line ratios measured in Perseus A \citep[top row; ][]{hatch06}, nuclei (middle row) and filaments (bottom row) for the 9 clusters in our sample. Note the overlap between the measured line ratios in Perseus A and in a typical filament.}
\label{lineratio_hist}
\end{figure}

In Figure \ref{compare_n2h} we compare our measured [\ion{N}{2}]/H$\alpha$ line ratios to those from the literature in order to assess the quality of these measurements. There is generally good agreement between this work and the literature, suggesting that our measured ratios are reliable to within $\sim$20\%. Much of this scatter is likely due to different choices of aperture and slit orientation, and likely does not represent the absolute uncertainty in our measurements. In Figure \ref{lineratio_hist} we compare our measured line ratios in nuclei and filaments to those measured in Perseus A \citep{hatch06}. In general, the line ratios measured at 98 different positions along the filaments in Perseus A agree with our measurements for the filaments in all 9 clusters. Furthermore, we find that the optical line ratios measured in Centaurus A \citep{farage10} agree with both Perseus A and our sample. Overall, the filaments in cool cores tend to have elevated low-ionization ratios, with the [\ion{O}{1}]/H$\alpha$ ratio being very narrowly peaked at $\sim$ 0.1. These line ratios are reminiscent of LINERs, but are measured far from the AGN, therefore suggesting a different origin.

%% - GAS KINEMATICS -- %%
\subsection{Gas Kinematics}

% Figure with ~2-D Spectra
\begin{figure*}[p]
\centering
\includegraphics[width=0.99\textwidth]{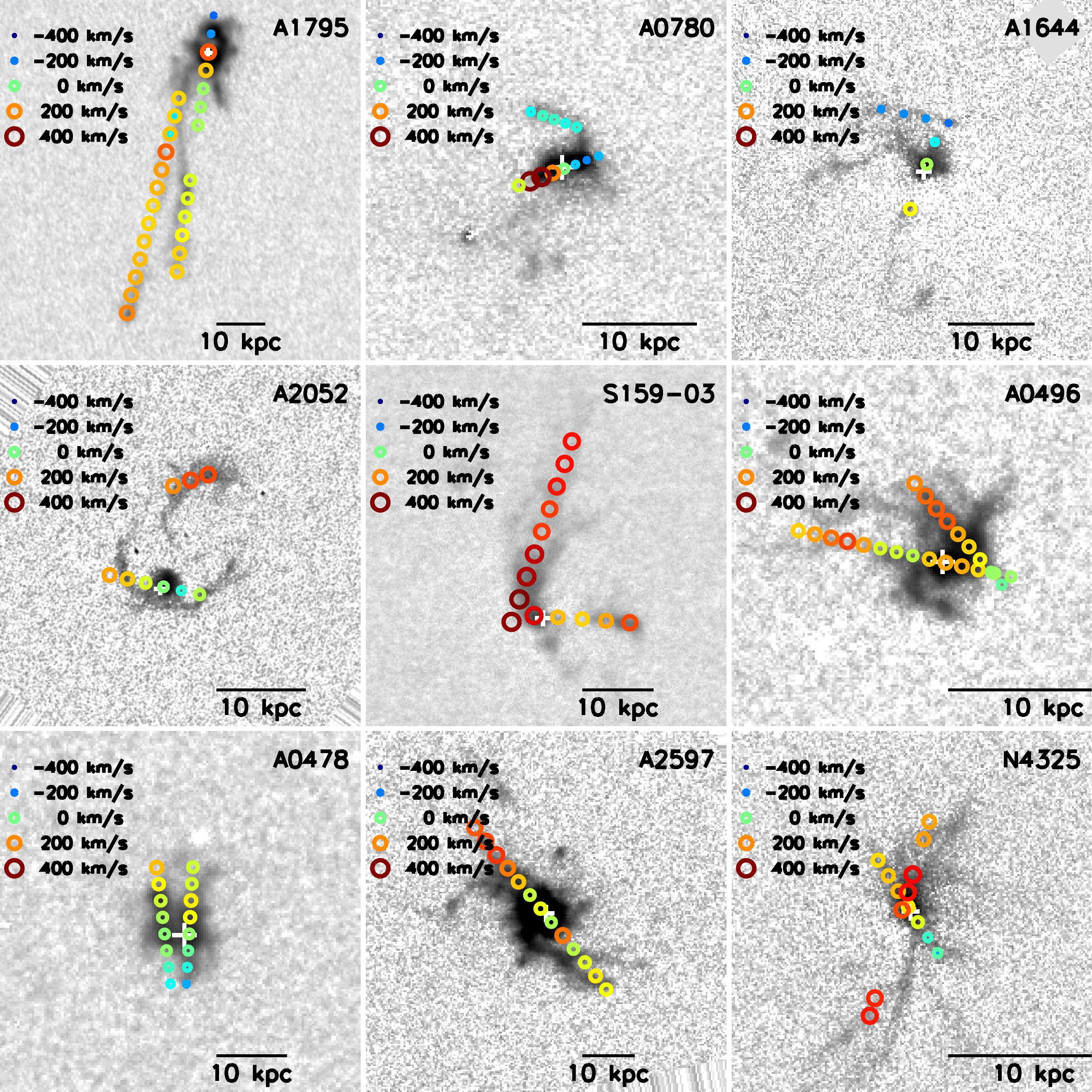}
\caption{Similar to figure \ref{3x3_n2h}, but showing the line-of-sight velocity of the optical line-emitting gas relative to the brightest cluster galaxy. This figure shows that the more extended filaments (Abell~0496, Abell~1795, Abell~2597, Sersic~159-03) tend to have relatively smooth velocity fields with modest velocity gradients.}
%\vspace{0.3in}
\label{3x3_vel}
\end{figure*}

\begin{figure}[p]
\centering
\includegraphics[width=0.8\textwidth]{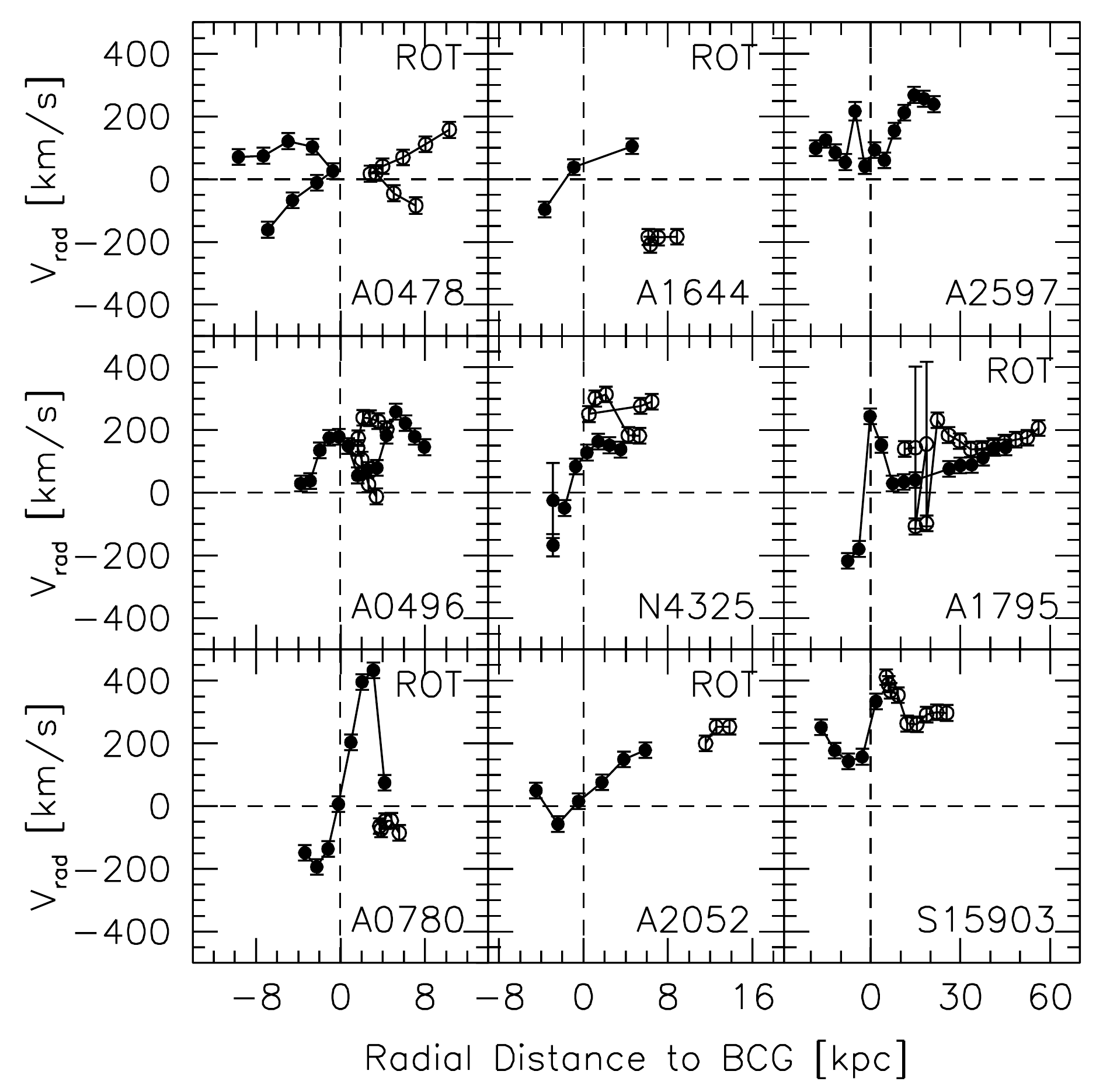}
\caption{Line-of-sight velocity of the optically-emitting gas relative to that of the BCG for 9 clusters in our sample. Different slit positions are differentiated by point type, and adjacent spatial pixels are joined by solid lines. At small radii (r $<$ 10kpc), there is evidence for rotation in several systems (Abell~0478, Abell~0780, Abell~1644, Abell~1795, Abell~2052) based on the measured $v/\sigma$ \citep{baum92}. Negative and positive radii correspond to positions along the slit with RA less than and greater than the BCG center, respectively.}
\label{velgrad}
\end{figure}

In Figure \ref{3x3_vel} we show H$\alpha$ maps for each cluster core, with the line-of-sight velocity measurements overlaid. These velocities represent the line-of-sight velocity difference between the warm gas and the central (BCG) stellar component. We find that, in general, systems with extended emission tend to have relatively smooth velocity fields (e.g., Abell~0496, Abell~1795, Abell~2597, Sersic~159-03), while compact systems and nuclei tend to have rotation signatures (e.g. Abell~0780, Abell~1644, Abell~0478). This observation is consistent with the work of \cite{baum90} and \cite{baum92} who found that emission-line nebulae in radio galaxies can be divided into three kinematic classes: rotators, calm nonrotators, and violent nonrotators. The systems which are classified as nonrotators (both calm and violent) tend to have relatively constant velocity fields and resemble the filaments in Figure \ref{3x3_vel}, while the rotators have strongly varying velocity fields, much like the nuclei and compact systems in Figure \ref{3x3_vel}. This is more easily seen in Figure \ref{velgrad}, which shows the line-of-sight velocity fields. In the central regions ($<10$ kpc) the velocity fields in many systems show evidence for rotation, with a characteristic shape reminiscent of a rotating disk. At large radius ($>10$ kpc), the velocity fields of the filaments appear to be relatively flat with typical line-of-sight velocities of $\sim$300 km s$^{-1}$ relative to the BCG. We do not find a correspondence between the kinematic class (rotator or non-rotator) as defined by \cite{baum90, baum92} and the ionization class (star-forming or shock-heated) described in \S3.1 and \cite{mcdonald11b}. 
We will discuss the implications of these kinematics further in \S4.2.

% Figure with ~2-D Spectra
\begin{figure*}[p]
\centering
\includegraphics[width=0.99\textwidth]{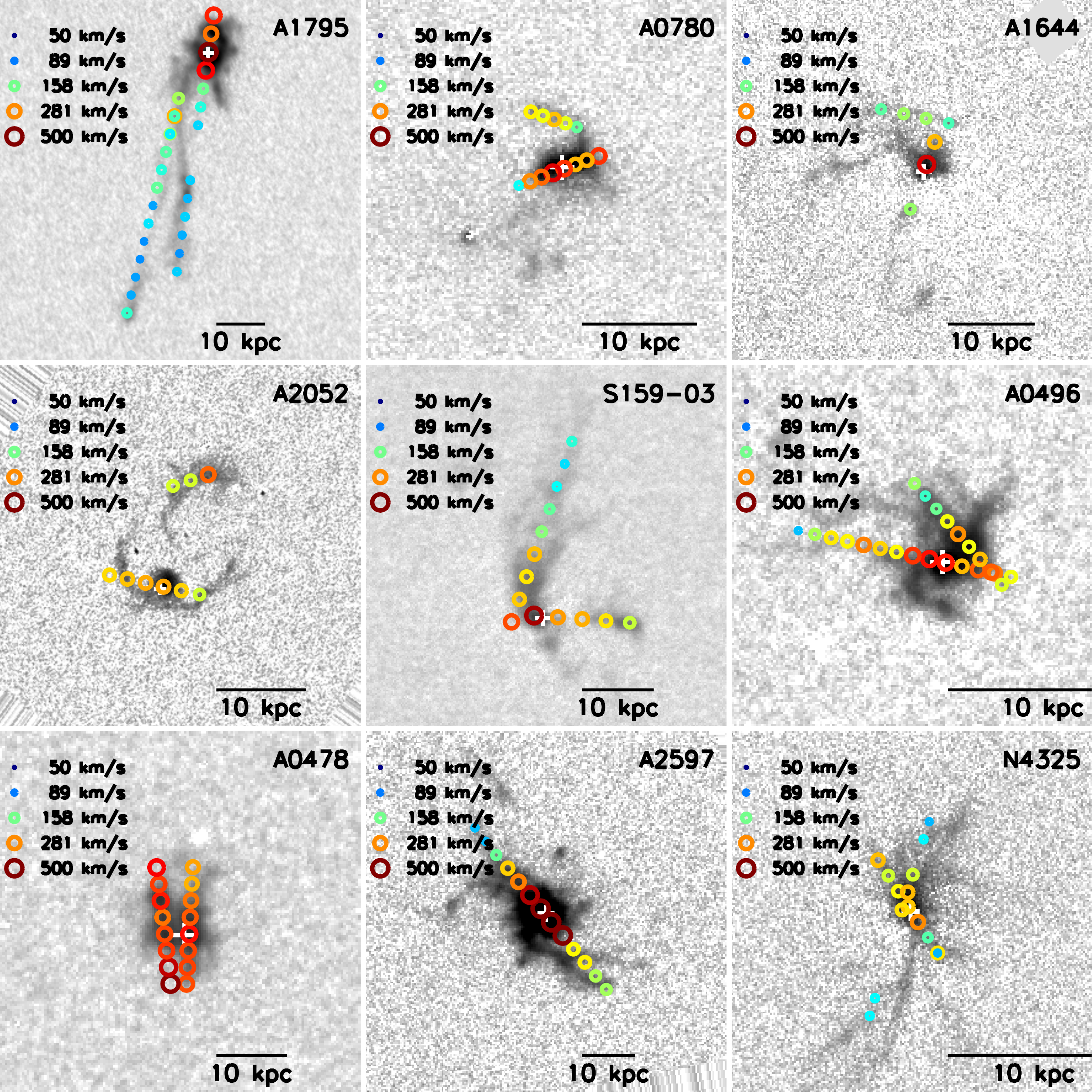}
\caption{Similar to Figure \ref{3x3_n2h} but now showing the velocity width. In nearly all systems the velocity dispersion peaks at the position of the optical nucleus. In the more extended systems there is a trend towards decreasing velocity dispersion with increasing radius, suggesting that turbulence is larger in the cool cores.}
\label{3x3_sig}
\end{figure*}

\begin{figure}[htb]
\centering
\begin{tabular}{c c}
\includegraphics[width=0.50\textwidth]{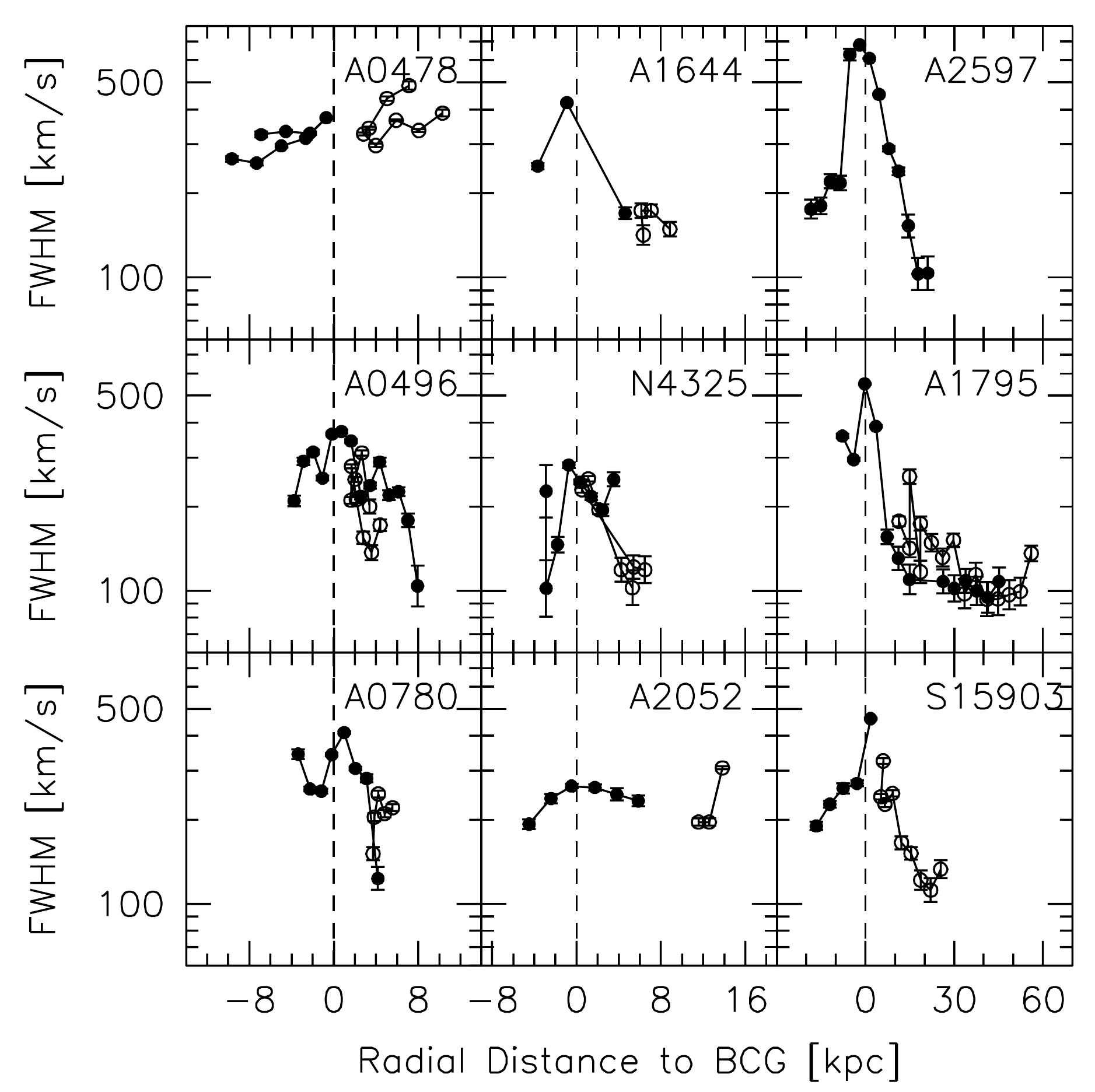} &
\includegraphics[width=0.46\textwidth]{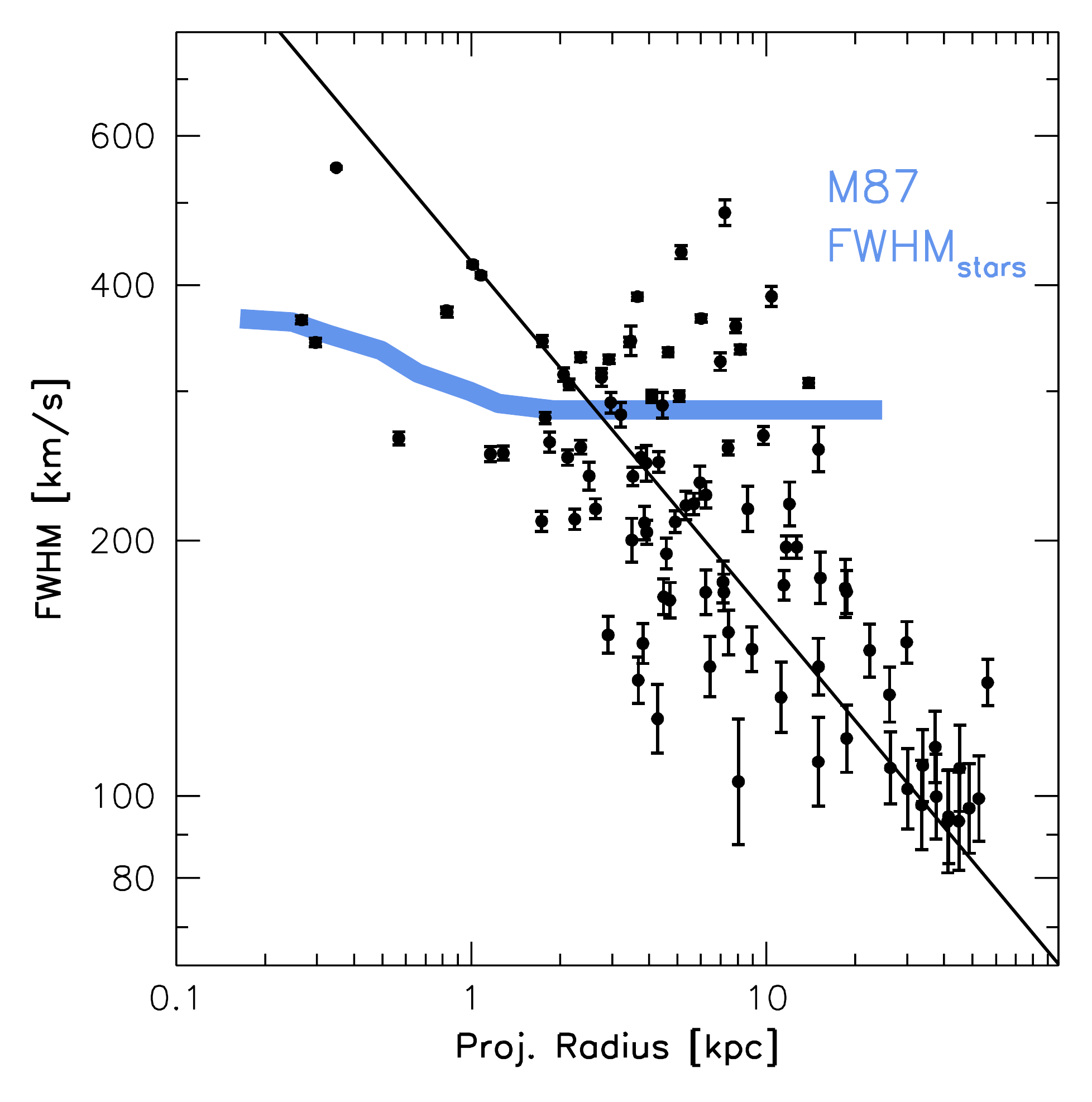} \\
\end{tabular}
\caption{Left: Similar to figure \ref{velgrad}, but now showing optical linewidth as a function of radius for individual clusters. Right: Optical linewidth as a function of radius for all 9 galaxy clusters. There is a strong correlation, such that the warm gas in the very center of the cluster is highly turbulent (FWHM $\sim 400$~km~s$^{-1}$) while the thin, extended filaments have very narrow lines (FWHM $< 200$~km~s$^{-1}$). This trend suggests that highly-elongated filaments can only survive in regions with minimal turbulence. Shown in blue is the stellar absorption line width in M87 from \cite{murphy11}. }
\label{sigr}
\end{figure}

In Figure \ref{3x3_sig} we show the velocity width of the filaments. In general, the velocity width peaks in the nucleus (FWHM $\sim 400$ km s$^{-1}$) while, in the filaments, it is consistently lower (FWHM $\sim 100-200$ km s$^{-1}$). The longest filaments (i.e., Abell~1795, Abell~2597, Sersic~159-03, NGC~4325) tend to have the narrowest emission lines. Curiously, Abell~2052, which harbors the most radio-luminous central AGN in our sample has the lowest central velocity width. Overall, the velocity width correlates well with the amount of turbulence one would infer from a visual inspection: long, thin filaments (e.g., Abell~1795, Abell~2597, Sersic~159-03) tend to have narrow emission lines, while short stubby filaments (e.g., Abell~0478) and regions with ``disturbed'' morphologies (e.g., Abell~0496, Abell~0780) tend to have broader profiles.

Figure \ref{sigr} shows the variation of the velocity width as a function of projected radius for each spatial element shown in Figure \ref{3x3_sig}. There is an obvious correlation (Pearson $R=-0.63$) between the optical linewidth and radial extent of the filaments. This trend was also reported, albeit with lower significance, by \cite{baum90} and \cite{baum92b}. This figure shows compelling evidence that the most extended optical filaments are also those which are experiencing the least amount of turbulence. If we assume that the H$\alpha$-emitting gas is a product of the ICM cooling \citep[e.g., ][]{mcdonald10}, this may tell us something about the amount of turbulence in the ICM within the cooling radius. Beyond the inner 10kpc, the optical linewidth is $<200$~km~s$^{-1}$, which is consistent with the observed upper limits on the ICM turbulence in cool cores from the \emph{XMM Reflection Grating Spectrometer} \citep[$\sim200$~km~s$^{-1}$;][]{sanders11}.  We note that the instrumental FWHM is $\sim 75$~km~s$^{-1}$, so our measurements are well above the lower limits.

The fact that we see broader lines near the center of the cluster could also be due to the higher likelihood of chance superposition of filaments at smaller radii, where the number density of filaments is higher. While the kinematic data alone are unable to differentiate between a true increase in velocity dispersion and a projection of multiple narrow lines, we are able to offer evidence for the former by looking at the optical line ratios. In \S3.1, we showed that the [\ion{N}{2}]/H$\alpha$ ratio is typically peaked in the nucleus, while in \S4.1 we will show further that there is a correlation between the velocity width and the low ionization line ratios. These trends suggest that the broad line widths that are observed at small radii are real and are not strongly affected by projection.

\begin{figure*}[htb]
\centering
\includegraphics[width=0.99\textwidth]{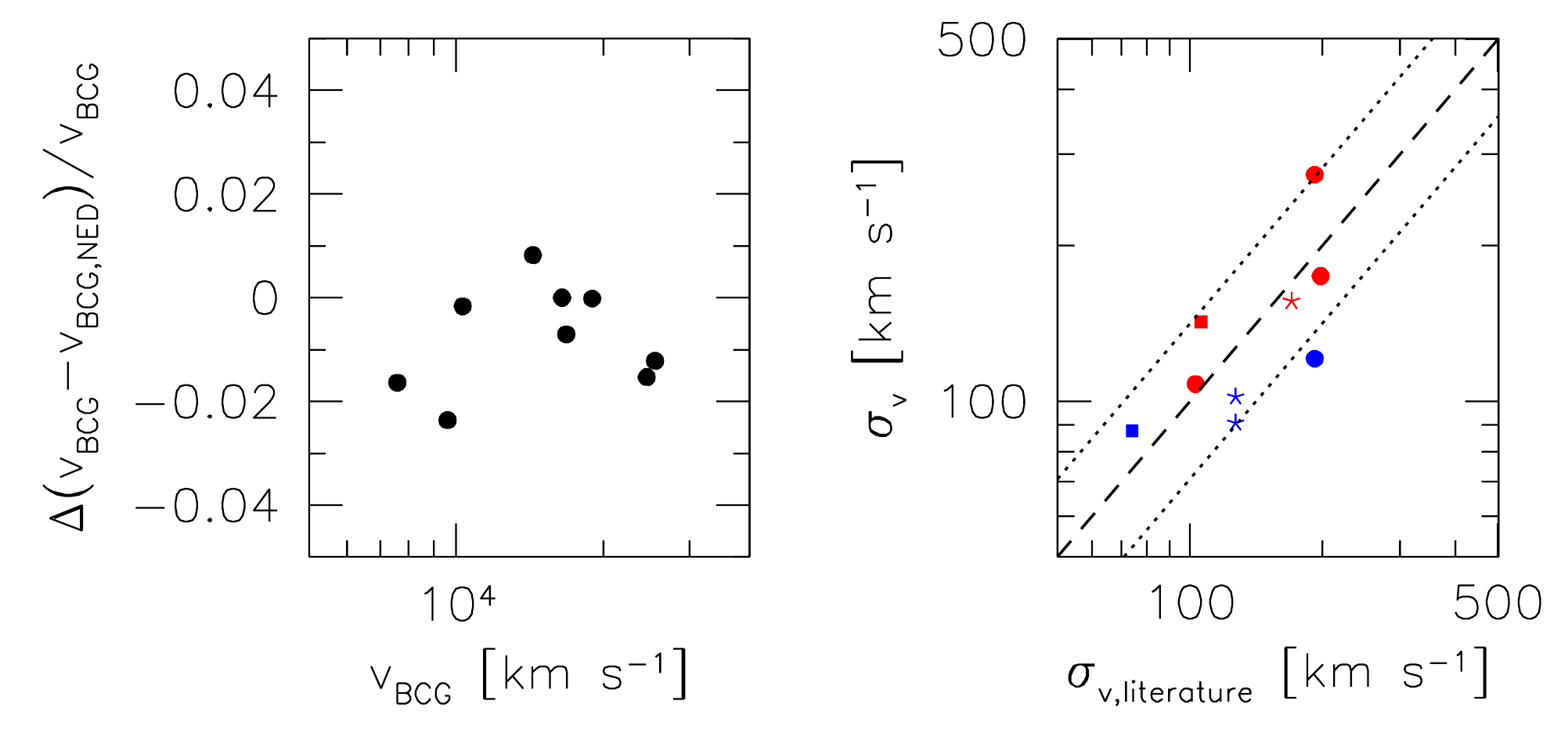}
\caption{Left: Comparison of BCG radial velocity measured for this work to those from the NASA Extragalactic Database for the same galaxies. This implies a typical error on our absolute velocity calibration of 1--2\%.  Right: Comparison of the velocity dispersion in the line-emitting gas for well-studied systems in the literature. Blue and red points represent extended and nuclear emission, respectively. Point types correspond to different references: \cite{heckman89} -- filled circles, \cite{melnick97} -- filled squares, \cite{hatch07} -- stars. The dashed line represents equality, while the dotted lines represent a scatter of 35\%. The large deviations are mostly dominated by the non-standard size of the apertures used in different studies.}
\label{compare_vel}
\end{figure*}

In Figure \ref{compare_vel} we compare our measured radial velocities and velocity dispersions to those from the literature. We find typical deviations of 1--2\% and $\lesssim$30\%  in radial velocity and velocity dispersion, respectively. We note that the fractional error in relative velocities (i.e. velocity of gas with respect to the BCG) is much lower and is independent of our absolute velocity calibration. The larger deviations in the velocity dispersion measurements are most likely a result of differing apertures between this work and previous studies of the same sources. Overall, there is good agreement between our kinematic measurements and those from previous studies.

These data tell an interesting story about the kinematics of the warm gas in the cool cores of galaxy clusters. In general, we find that longer filaments are less turbulent than shorter or morphologically peculiar filaments. The velocity width in the warm gas increases with decreasing radius, reaching a peak in the nucleus of $\sim 400$~km~s$^{-1}$, and a minimum of $\sim 100$~km~s$^{-1}$ beyond 10 kpc. The velocity fields are typically well-ordered along the filaments. We return to these results in \S4.2 below.

\subsection{Intrinsic Reddening}
The amount of dust in the ICM is still uncertain. The abundance of relativistic particles should act to destroy dust grains on short timescales \citep{draine79}. However, the fact that we observe star formation and, more importantly, mid--far IR emission \citep[e.g., ][]{odea08}, suggests that some amount of dust may be shielded from destructive processes. Our data provide new estimates on the amount of intrinsic reddening in the optical filaments and nuclei in the BCGs of cool core clusters -- something that is currently lacking in the literature for more than a few systems \citep[e.g., ][]{hu85, crawford99}. In order to determine the amount of intrinsic reddening, we began with the H$\alpha$/H$\beta$ line ratio, corrected for Galactic extinction and underlying stellar absorption (\S2). We made the assumption that systems with H$\alpha$/H$\beta > 2.85$ (corresponding to case B recombination) had non-zero intrinsic reddening and determine the intrinsic $E(B-V)$ assuming a dust screen model and a standard extinction curve \citep[$R_v=3.1$; ][]{cardelli89}. The reddening estimates here assume a foreground screen model and, thus, are underestimated by a factor of $[\exp(\tau)-1]/\tau$ if we assume that the line-emitting gas and dust are well-mixed.

% Figure with ~2-D Spectra
\begin{figure*}[p]
\centering
\includegraphics[width=0.99\textwidth]{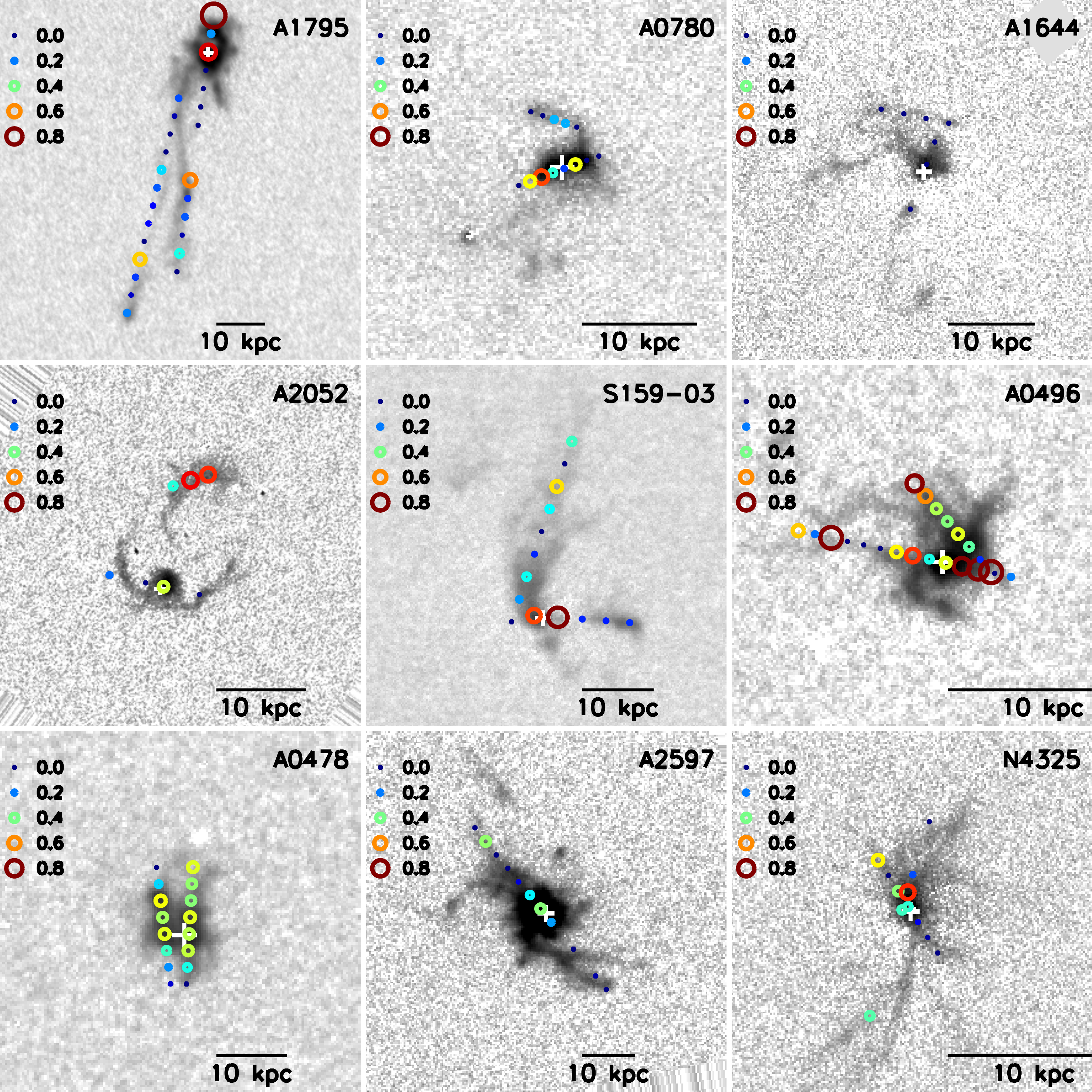}
\caption{Similar to Figure \ref{3x3_n2h} but now showing the intrinsic reddening, E(B-V), corrected for Galactic extinction. We note that many of the regions with locally high E(B-V) are coincident with low H$\alpha$ surface brightness (e.g., Abell~0496, Abell~1795, Sersic~159-03) suggesting that they are more uncertain.}
\label{3x3_ha2hb}
\end{figure*}

In Figure \ref{3x3_ha2hb} we provide the 2-D distribution of reddening estimates based on ($E(B-V)$), corrected for Galactic extinction, for the 9 systems in our sample. Typical uncertainties in these estimates, assuming that the $\sim$10--15\% uncertainty in f$_{H\beta}$ dominates, is $\Delta E(B-V) \sim 0.15$. With the exception of a few outliers, the vast majority of systems appear to be mostly free of intrinsic reddening. The five most extended systems, Abell~1644, Abell~1795, Abell~2597, NGC~4325, and Sersic~159-03 all appear nearly reddening-free. There is a significant amount of reddening in the core of Abell~0780 (Hydra A), which is known to have a dust-lane at approximately the position of the reddening peak. Abell~0478, which has the highest Galactic $E(B-V)$, seems to have a moderate amount of intrinsic reddening everywhere, but we cannot formally rule out a Galactic origin due to small fractional uncertainty on the large Galactic extinction value. Both Abell~0496 and Abell~2052 appear to have legitimate, non-zero reddening in the filaments. In the case of Abell~2052, the reddened filaments (north of the BCG nucleus) appear to be shock-heated (McDonald \etal 2011b) based on the high H$\alpha$/FUV ratio. 

\begin{figure}[h!]
\centering
\includegraphics[width=0.7\textwidth]{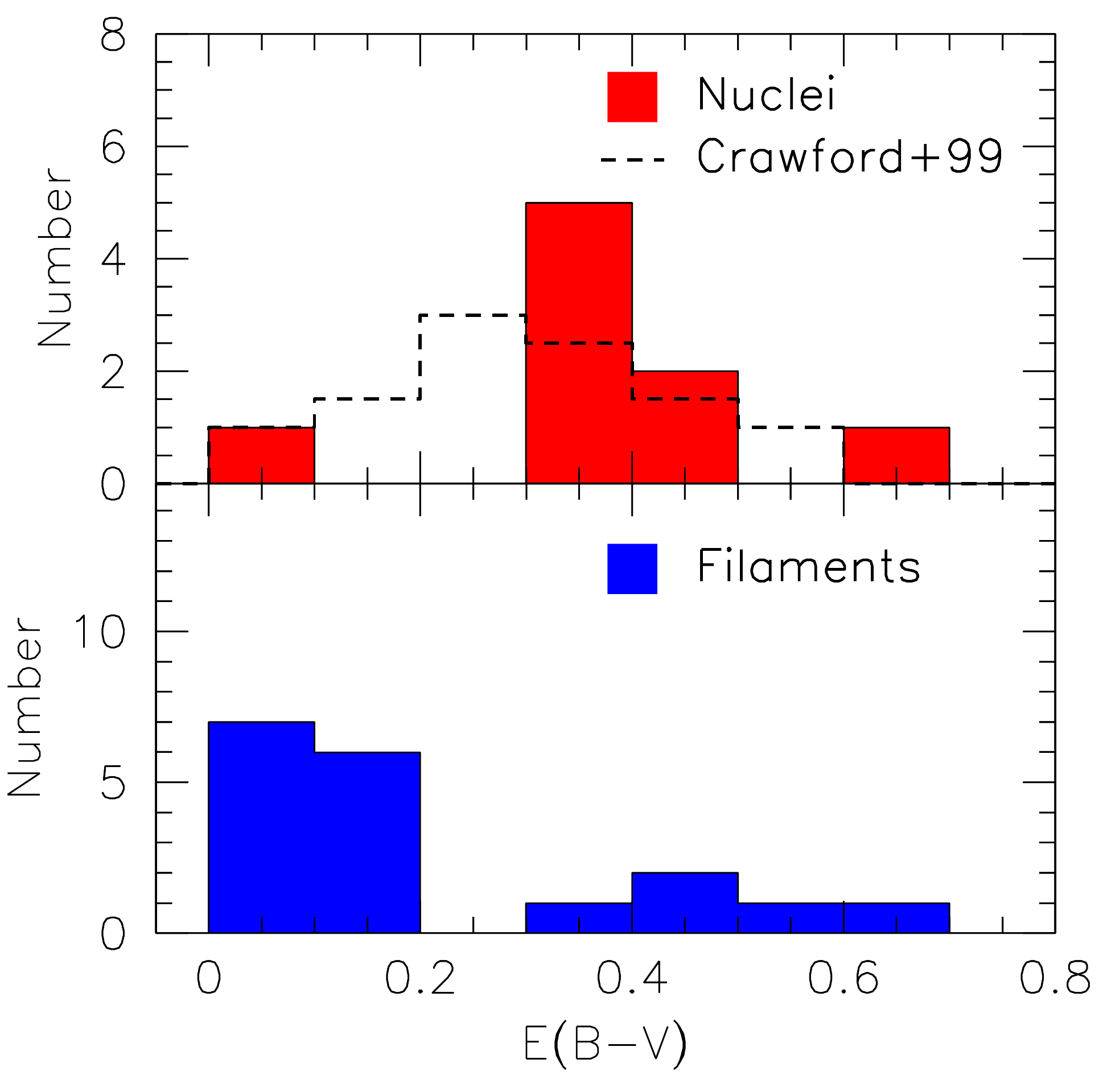}
\caption{Distribution of optical reddening, $E(B-V)$, in the nuclei (upper) and filaments (lower) of 9 cool core clusters. The amount of reddening in nuclei is relatively flat over the range $0 < E(B-V) < 0.4$, while, in the filaments, the distribution peaks sharply at $E(B-V)=0$, with a tail out to $E(B-V) \sim 0.6$. The distribution of $E(B-V)$ in the nuclei matches well with that found by \cite{crawford99} (dotted line). The fact that the majority of filaments have little-to-no reddening suggests that dust does not survive long in the intracluster medium. The objects with $E(B-V)>0.4$ are Abell~0478, Abell~0496, and Abell~2052, as discussed in the text. These reddening estimates assume a foreground screen model and, thus, are underestimated by a factor of $[\exp(\tau)-1]/\tau$ if we assume that the line-emitting gas and dust are well-mixed.}
\label{ha2hb}
\end{figure}

We can also consider the distribution of reddening measurements in filaments and nuclei. Using the same regions as in Figure \ref{lineratios}, we take the emission-weighted average $E(B-V)$ in each region. The distribution of these values is shown in Figure \ref{ha2hb}. For the nuclei there is a relatively broad distribution in $E(B-V)$ from 0.0--0.7, while in filaments the distribution peaks at $E(B-V)=0$ with a broad tail out to $E(B-V)=0.6$. We find that $\sim$70\% of filaments have $E(B-V)<0.2$, contrary to previous estimates based on randomly-oriented slits (e.g., Crawford \etal 1999) which found considerable amounts of optical reddening in cool core clusters. However, previous measurements contained only the nuclear emission and, thus, a comparison to our in-filament reddening estimates is unfair. We find good agreement between this work and \cite{crawford99} for the distribution of reddening in nuclei, as shown in Figure \ref{ha2hb}. The consistently small amount of reddening derived in the H$\alpha$ filaments appears to be inconsistent with the evidence in the UV and mid-infrared of on-going star formation \citep[e.g., ][]{rafferty06, odea08, hicks10, mcdonald11b}, but is consistent with previous findings for the warm gas in Perseus A \citep[E(B-V)$\sim$0.2;][]{sabra00} and Centaurus A \citep[E(B-V)=0.0--0.7;][]{canning11}.  We speculate that the harsh conditions surrounding the ionized filaments may prevent dust from surviving in all but the densest regions \citep{draine79}. Finally, we note that these results are robust to our choice of H$\alpha$/H$\beta=2.85$ as the dust-free limit and do not change dramatically if we instead choose H$\alpha$/H$\beta=3.1$, which reflects the conditions in shock-heated gas.

%================================================================%
%============== DISCUSSION ======================================%
%================================================================%

\section{Discussion}
The data presented in \S3 represent a significant step forward in combined spatial and spectral resolution and coverage over prior large spectroscopic surveys of emission-line nebulae in cool core clusters \citep[e.g., ][]{heckman89, baum90, allen92, crawford99, hatch07, edwards09}. Such a dataset allows us to provide new constraints on the origin and ionization mechanism in these filaments. In the following subsections we discuss the implications of the results presented thus far and elaborate on their meaning. 

\subsection{Sources of Ionization}
There are several possible sources of ionization in the cool cores of galaxy clusters. The most popular ideas have been: (a) the central AGN, (b) young stellar populations, (c) X-rays from the ICM, (d) heat conduction from the ICM to the cold filament, (e) shocks and turbulent mixing layers, and (f) collisional heating by cosmic rays. The lack of a strong gradient in [\ion{N}{2}]/H$\alpha$ with radius (Figure \ref{3x3_n2h}) and the low [\ion{O}{3}]/H$\beta$ ratio in the filaments suggests that AGN do not contribute significantly to the ionization outside of the nucleus.  The relative weakness of high-ionization lines (i.e., [\ion{O}{3}]) suggests that ionization by ICM X-rays is also a small contributor. While scenarios (a) and (c) are relatively easy to rule out in the filaments, the remaining scenarios require a more quantitative approach.

In Figure \ref{models1} we show the same data as in Figure \ref{lineratios}, but now include model expectations for star formation, cooling plasma, shocks, conduction, and collisional ionization by cosmic rays. The model grids representing photoionization by young stars \citep[upper panels; ][]{kewley01} show partial overlap with the data for filaments in the [\ion{N}{2}]/H$\alpha$ and [\ion{S}{2}]/H$\alpha$ panels. However, these models fail to produce adequately high [\ion{O}{1}]/H$\alpha$ ratios, as noted by earlier studies \citep[e.g., ][]{crawford99}.  The best match between the model and data is achieved if we use a model cloud with roughly solar metallicity and a low total ionization parameter (U). The location of this grid would move slightly if the IMF was altered, but no reasonable combination of ionization parameter, metallicity, and IMF would reproduce the observed [\ion{O}{1}]/H$\alpha$ ratios. 

\begin{figure*}[p]
\centering
\includegraphics[width=0.99\textwidth]{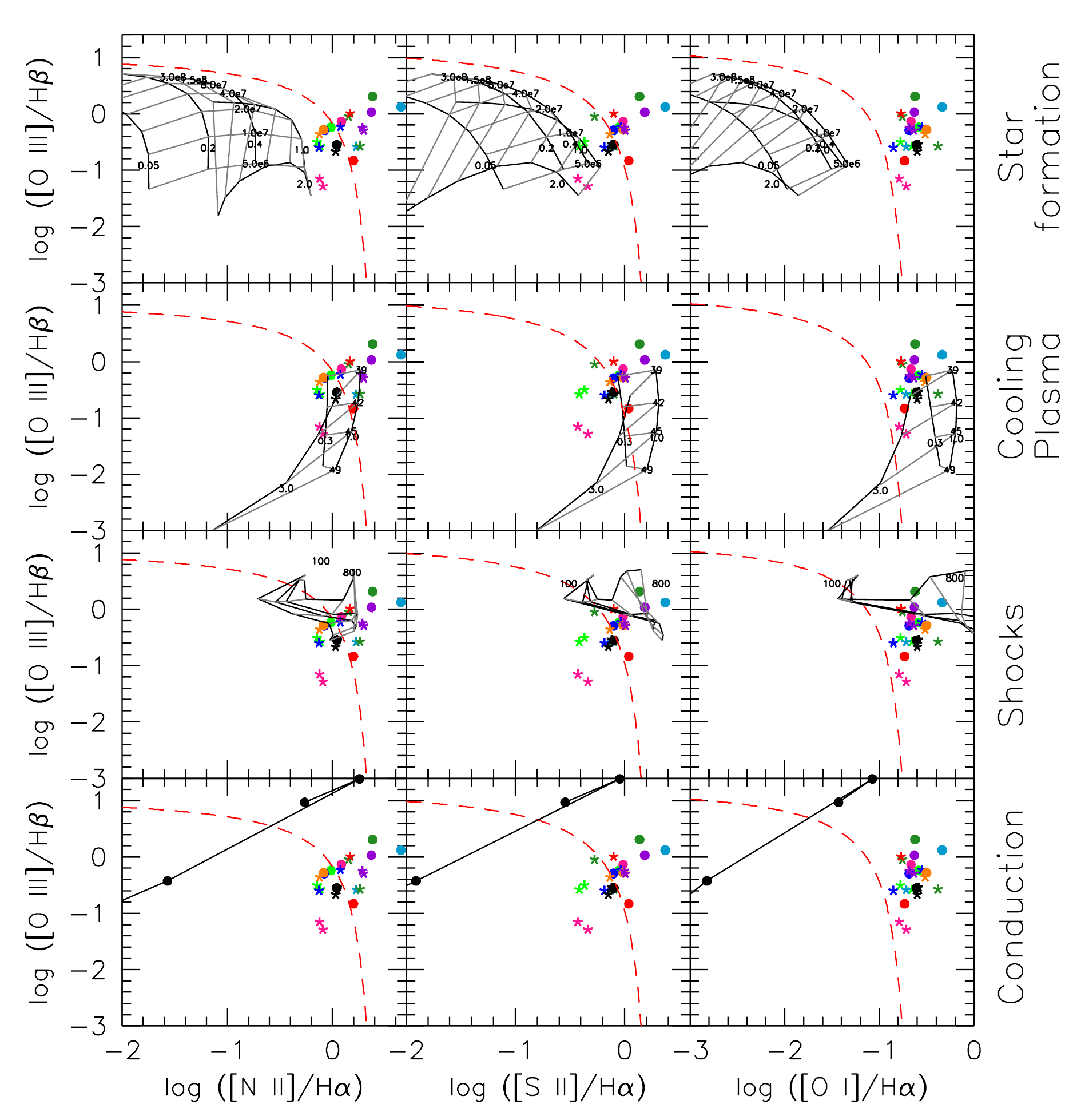}
\caption{Similar to Figure \ref{lineratios}, but with model predictions overlaid. From top to bottom: photoionization from young stars \citep[lines of constant ionization parameter and metallicity; ][]{kewley01}, self-ionization from a condensing plasma \citep[lines of constant ionization parameter and metallicity; ][]{voit94}, fast shocks \citep[lines of constant speed and magnetic field strength; ][]{allen08},  conduction \citep[points correspond to different initial conditions; ][]{boehringer89}, and collisional ionization by cosmic rays \citep{ferland09}. In all panels, the point/color types are consistent with those in Figure \ref{lineratios} and the red dashed line is the extreme starburst limit of \cite{kewley06}.}
\label{models1}
\end{figure*}

\addtocounter{figure}{-1}
\begin{figure*}[h!]
\centering
\includegraphics[width=0.99\textwidth]{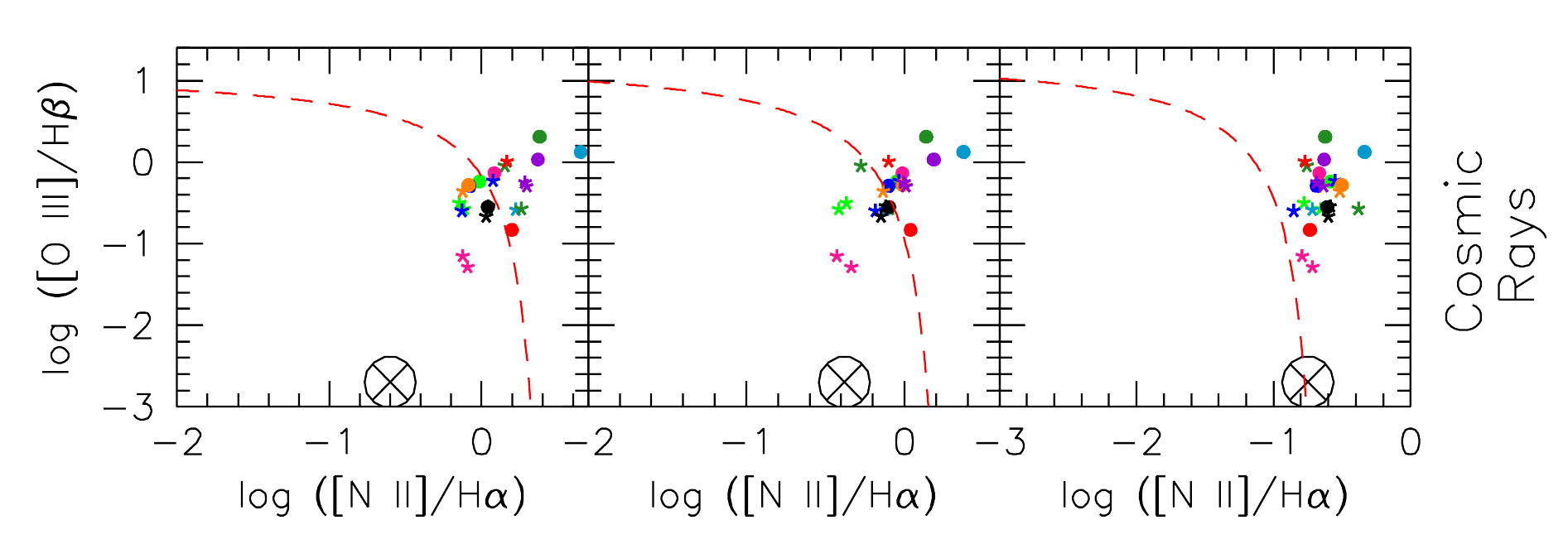}
\caption{Continued.}
\end{figure*}

\begin{figure}[htb]
\centering
\includegraphics[width=0.6\textwidth]{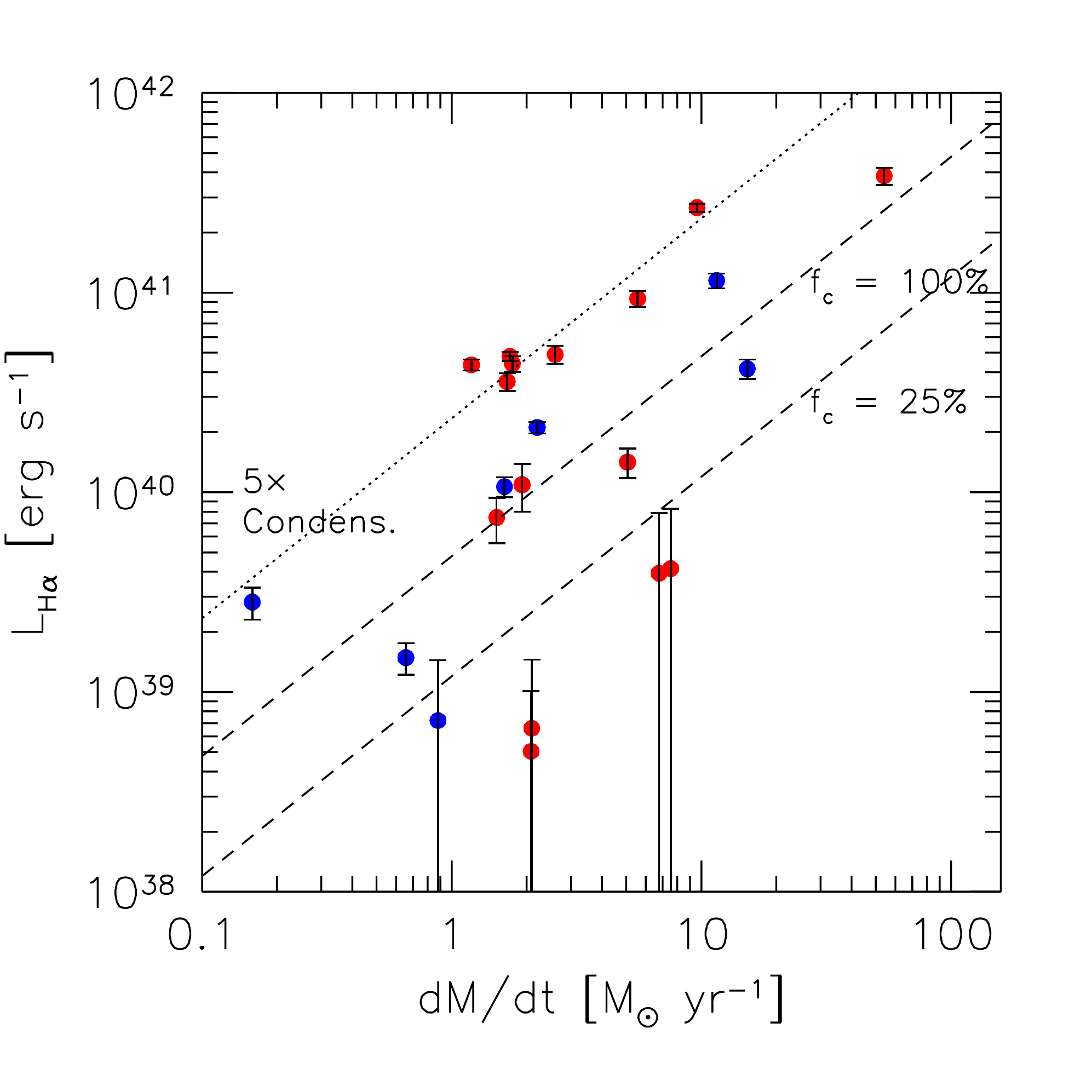}
\caption{Total H$\alpha$ luminosity versus X-ray-derived cooling rates from \cite{mcdonald10} and \cite{mcdonald11a}. The red and blue points correspond to galaxy clusters and groups, respectively. The dashed lines show the predicted H$\alpha$ luminosity assuming that the flux is due to self-ionization by a condensing hot plasma, for covering factors of 25\% and 100\%. The dotted line is five times the expected H$\alpha$ luminosity for a covering factor of 100\%. For the most extended and luminous filaments, this process contributes, at most, 20\% of the H$\alpha$ flux, assuming an unrealistically high covering factor.}
\label{dmdt}
\end{figure}

In the second row of Figure \ref{models1}, we show the expectation for condensing intracluster gas, originating at $10^7$~K \citep{voit94}. These models assume that the intracluster gas at $T>10^4$~K photoionizes the $10^4$~K gas as it cools radiatively. In choosing the models to plot we make the assumption that the cool gas is ionization-bounded. This assumption is based on the fact that we observe strong [\ion{O}{1}] and [\ion{S}{2}] lines -- species that are typically observed beyond the classical H~II region \citep{hatch06}. These models reproduce well the observed [\ion{N}{2}]/H$\alpha$ and [\ion{O}{3}]/H$\alpha$ ratios observed in the filaments, but tend to slightly over-predict the [\ion{S}{2}]/H$\alpha$ and [\ion{O}{1}]/H$\alpha$ ratios. Interestingly, these models err in the opposite direction as the models for star formation, suggesting that a combination of the two may provide an adequate fit to the data. An important test for this model is whether the observed cooling rates can produce the high optical line luminosities that we observe. In \cite{donahue91}, the total H$\alpha$ luminosity due to the condensing ISM is given as:

\begin{equation}
\left(\frac{L_{H\alpha}}{10^{41} \rm{~erg~s^{-1}}}\right) = \left(\frac{\dot{M}_{H\alpha}}{16 \rm{~M_{\odot} ~yr^{-1}}}\right)(T_7f_c\epsilon_3)
\label{cflow}
\end{equation}

where $T_7$ is the maximum temperature in units of $10^7$~K, $f_c$ is the fraction of the cooling radiation incident upon the photoionized clouds, and $\epsilon_3$ is the fraction of incident radiation that reemerges in the H$\alpha$ line normalized to 3\%. We use $\epsilon_3 = 0.77$, from \cite{voit94}. In Figure \ref{dmdt} we show the H$\alpha$ luminosity as a function of the X-ray-derived cooling rates for 14 clusters (red) from \citet{mcdonald10} and 7 groups (blue) from \citet{mcdonald11a}. Even if we make the unrealistic assumption that 100\% of the cooling luminosity is incident upon the cool clouds ($f_c = 1$), the majority of systems are still much more luminous at H$\alpha$ than predicted. Under these extreme conditions, the most luminous systems are consistent with still only $<$ 20\% of their H$\alpha$ emission coming from the condensing ICM, in agreement with earlier results by \cite{voit90} using the same methods.  Thus, while the model of a self-radiating cooling flow is promising in terms of the predicted line ratios, the total fluxes are too low in the context of more recent, spectroscopically-derived X-ray cooling rates. However, the fact that we do measure non-zero cooling rates in the ICM means that self-radiation due to the cooling ICM is a contributing process and should not be ignored -- a point we will return to later.

\begin{figure}[p]
\centering
\includegraphics[height=0.8\textheight]{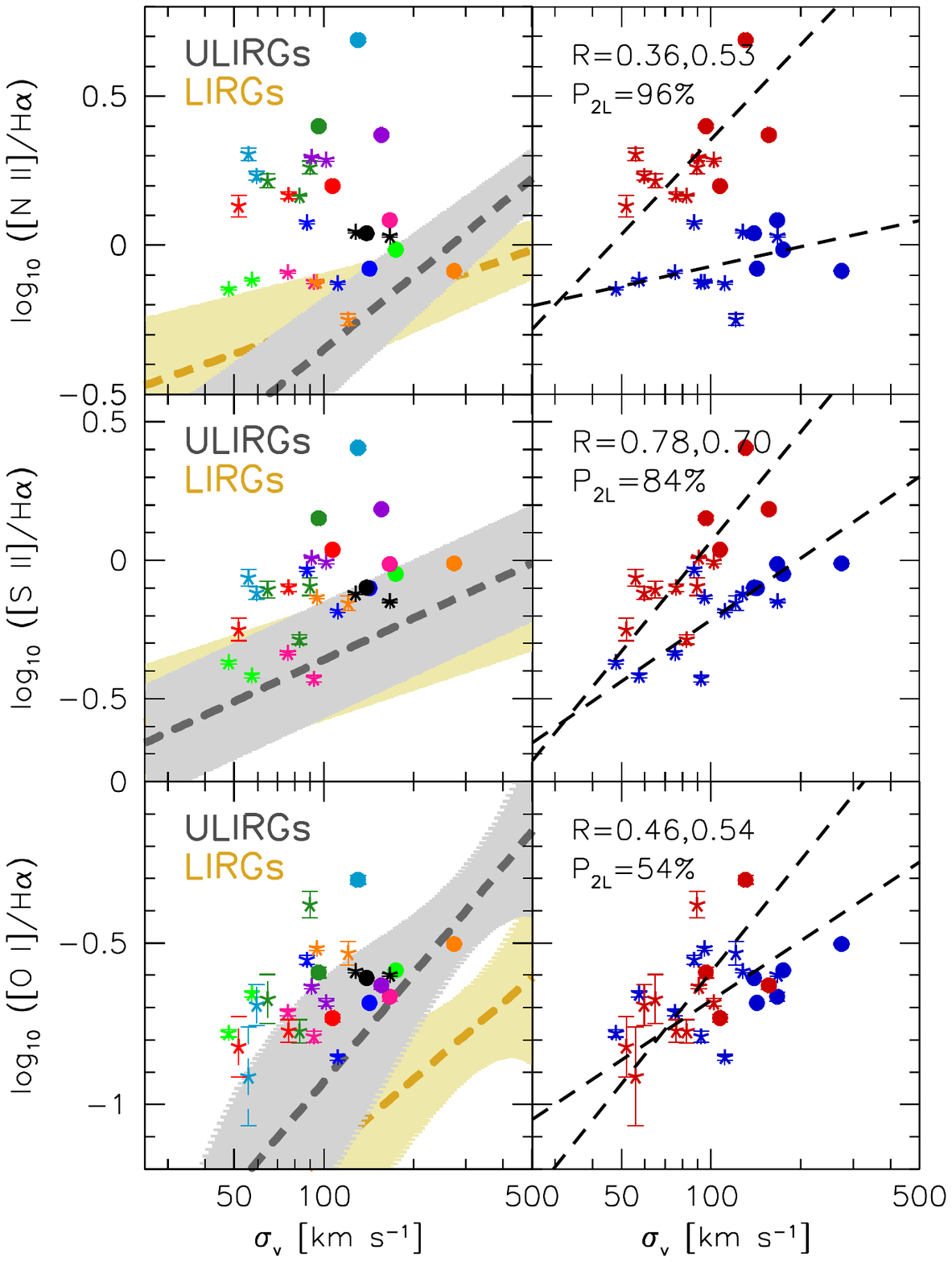}
\caption{[\ion{N}{2}]/H$\alpha$, [\ion{S}{2}]/H$\alpha$, and [\ion{O}{1}]/H$\alpha$ line ratios as a function of the velocity dispersion for nuclei and filaments in our sample. Point types/colors are described in Figure \ref{lineratios}. In the left panels, the regions occupied by LIRGs and ULIRGs \citep{monreal10} are shown in grey and yellow, respectively. In the right panels, we show two-line fits to the data, with the Pearson R of each line in the upper left corner. Also shown in the upper left corner is the F-test probability, $P_{2L}$, with which we can reject the null hypothesis of a single-line fit in favor of a two-line fit.}
\label{sig_lineratios}
\end{figure}

In the third row of Figure \ref{models1}, we show model expectations for shocks of various speeds and magnetic field strengths from \cite{allen08}. In general, the shock models cover the same range in [\ion{N}{2}]/H$\alpha$, [\ion{S}{2}]/H$\alpha$, and [\ion{O}{1}]/H$\alpha$ as the data, but slightly over-predict [\ion{O}{3}]/H$\beta$. The low-ionization ratios are best matched in the filaments by slow shocks ($v\sim 100-400~$km s$^{-1}$), and in the nuclei by fast shocks ($v\sim800~$km s$^{-1}$). This makes qualitative sense, since the line widths are considerably broader in the nucleus than in the filaments. To further investigate the role of shocks, we consider the velocity dispersion as a function of the [\ion{N}{2}]/H$\alpha$, [\ion{S}{2}]/H$\alpha$, and [\ion{O}{1}]/H$\alpha$ ratios in Figure \ref{sig_lineratios}. We find that the systems with the weakest low-ionization line ratios have velocity dispersions consistent with those found for luminous and ultraluminous infrared galaxies (hereafter LIRGs and ULIRGs, respectively), which are likely experiencing a mix of shocks and star formation \citep[][]{veilleux95, veilleux99, monreal10}. The systems which have low ionization line ratios considerably larger than those found in ULIRGs and LIRGs were identified earlier as non star forming based on the optical line ratios and far-UV emission. In the right panels of Figure \ref{sig_lineratios}, we show that these non star forming systems have  [\ion{N}{2}]/H$\alpha$ and [\ion{S}{2}]/H$\alpha$ ratios which are more strongly dependent on the velocity dispersion than their star forming counterparts. One possible interpretation of this result is that the filaments and nuclei are ionized by a mix of shocks and star formation. In this scenario, the two lines in Figure \ref{sig_lineratios} represent two extremes: the shock-dominated case (steep line) and star formation-dominated case (shallow line). We will further investigate this scenario in the next section.

In the fourth row of Figure \ref{models1} we show the expectation for heat conduction along the boundary between the hot ICM and the cool filaments from \cite{boehringer89}. This mechanism produces either too high [\ion{O}{3}]/H$\alpha$ ratio or too low [\ion{N}{2}], [\ion{S}{2}], and [\ion{O}{1}] ratios, depending on the choice of conditions. While the presence of soft X-ray emission coincident with optical filaments is evidence that conduction may be occurring \citep[e.g., ][]{nipoti04}, it cannot be the dominant source of ionization of the optical filaments on the basis of the observed optical line ratios.

Finally, the bottom panels of Figure \ref{models1} shows the predicted line ratios for collisional ionization by cosmic rays \citep{ferland09}. While this model accurately predicts the [\ion{S}{2}]/H$\alpha$ and [\ion{O}{1}]/H$\alpha$ ratios, it underpredicts the [\ion{N}{2}]/H$\alpha$ and [\ion{O}{3}]/H$\beta$ ratios by 1 and 3 orders of magnitude, respectively. In the absence of an additional ionization source which can produce strong [\ion{O}{3}] and [\ion{N}{2}] emission while producing very little H$\alpha$, H$\beta$, [\ion{S}{2}], and [\ion{O}{1}], these large discrepancies effectively rule out cosmic rays as a strongly contributing ionization mechanism. However, we point out that, while they do not appear to be a sufficient ionization mechanism for the cool gas, cosmic rays may still be important in providing energy to the ICM, heating $H_2$ \citep{ferland09, donahue11}, and helping to prevent massive cooling flows \citep[e.g., ][]{mathews09}.

It is clear from Figures \ref{models1}--\ref{sig_lineratios} that, individually, none of the models we listed at the beginning of this section can simultaneously explain all optical line ratios, their observed dependence on linewidths, and the total H$\alpha$ luminosity. However, there is a very real possibility that these filaments are, in fact, ionized by a combination of processes. Below, we discuss two such composite models which offer promising results.

% Cooling ICM - Voit 92
% Photoionization - Kewley
% Conduction - Bohringer 89
% Shocks - Allen 08
% Cosmic Rays - Ferland 09

\subsubsection{Composite Models -- Star Formation in a Condensing ICM}

There is compelling evidence for both star formation and cooling flows in the cool cores of many galaxy clusters. Modest cooling flows are observed in the X-rays \citep[e.g., ][]{peterson03, voigt04} and EUV \citep[e.g., ][]{oegerle01a, bregman06}, while the evidence for star formation is present at UV, optical, and IR wavelengths as we presented in \S3.2. Thus, a composite model including self-ionizing, cooling gas and star formation is both observationally motivated and straightforward to interpret. In order to test this model, we turn to Eq. \ref{cflow}, which predicts the H$\alpha$ luminosity given an ICM cooling rate. We make the assumption that the cooling ICM surrounds the cool clouds and, thus, only half of the ionizing radiation is incident on the cool gas. We then compute the expected H$\alpha$ luminosity on a cluster-by-cluster basis, using X-ray-derived cooling rates from \cite{mcdonald10} and \cite{mcdonald11a}. Finally, we compare the predicted H$\alpha$ luminosity to the measured values from the aforementioned references and determine the fraction of H$\alpha$ luminosity in each cluster that is due to the condensing ICM. We further assume that this fraction is constant throughout the cluster and remove the contribution to each emission line flux from the cooling flow, assuming the predicted line ratios shown in Figure \ref{models1}.

\begin{figure}[htb]
\centering
\begin{tabular}{c c c}
\includegraphics[width=0.32\textwidth]{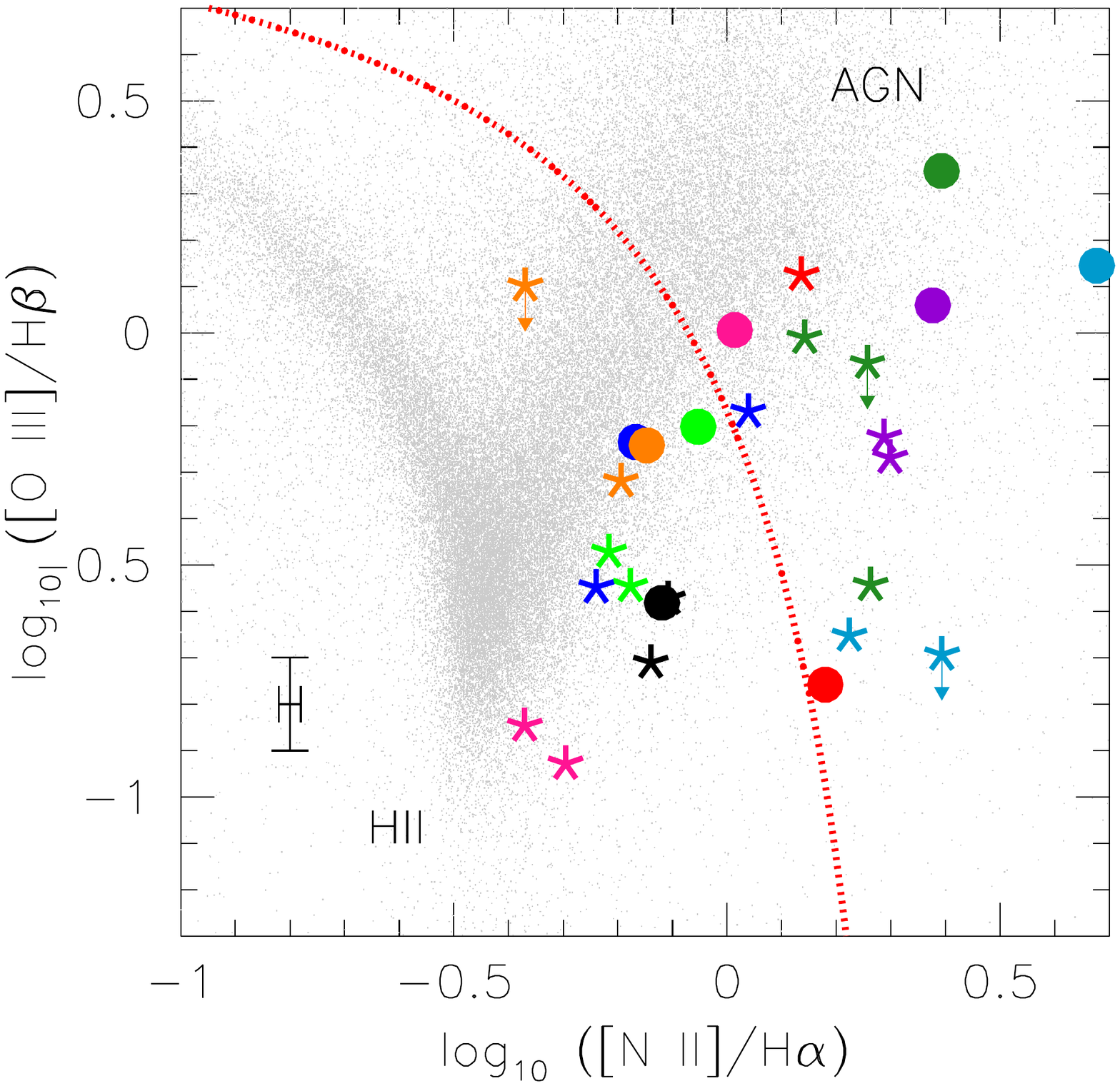} &
\includegraphics[width=0.32\textwidth]{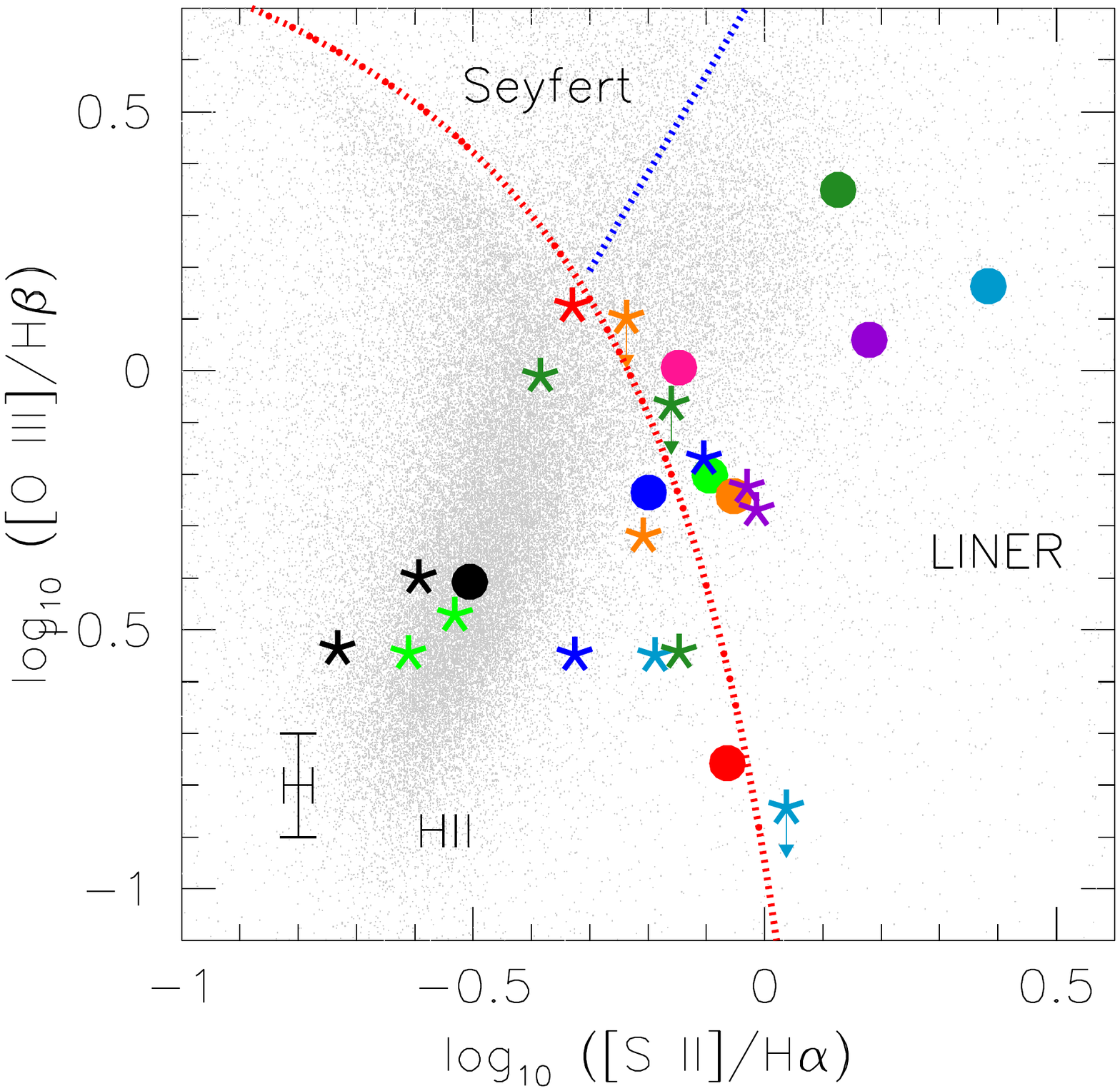} &
\includegraphics[width=0.32\textwidth]{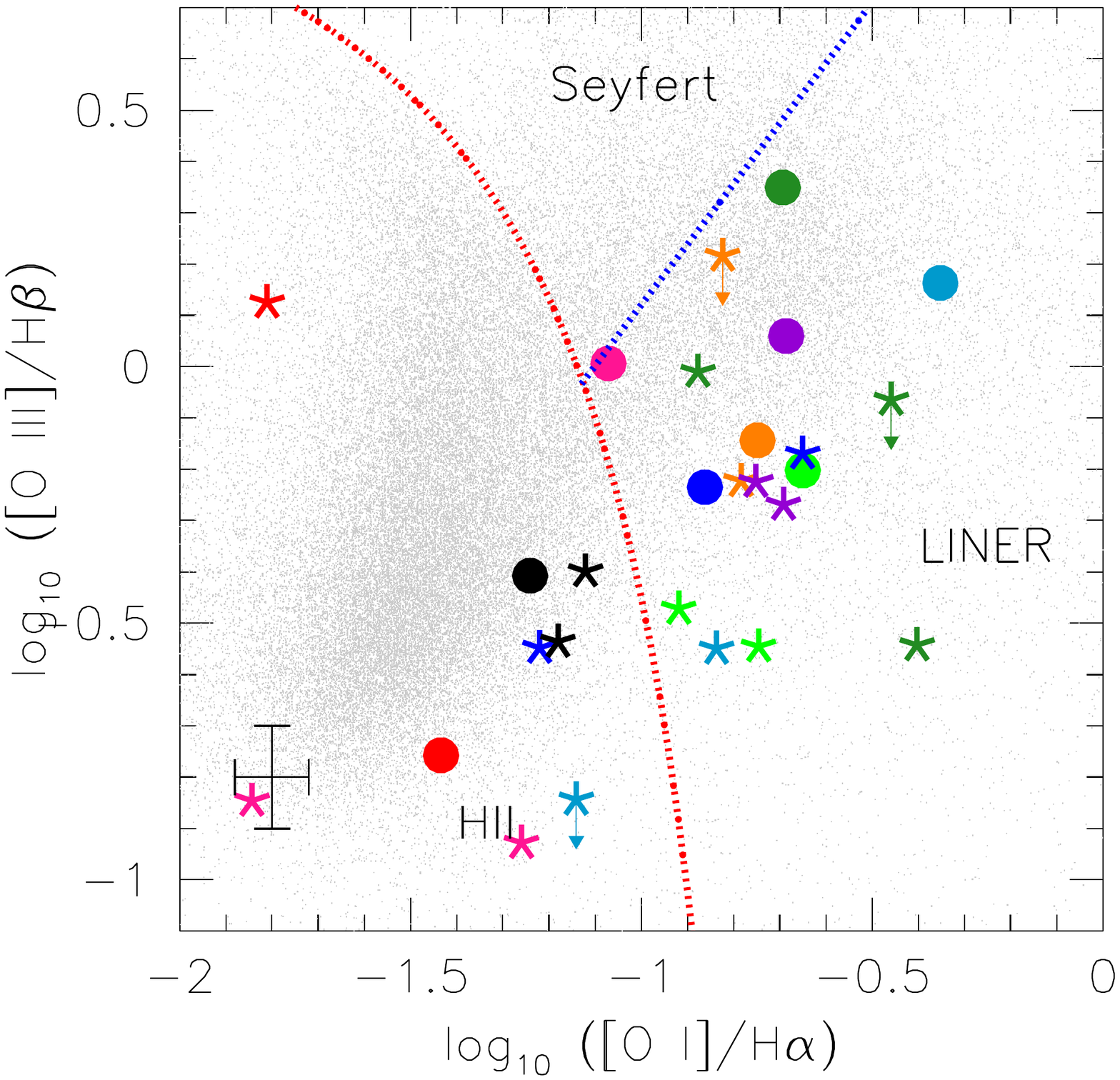} \\
\end{tabular}
\caption{Similar to figure \ref{lineratios}, but with the expected contribution from the cooling flow removed. By subtracting the expected flux based on the X-ray cooling properties of each cluster, we find a better match between the optical emission line ratios in cool core cluster filaments and typical \ion{H}{2} regions, with the exception of the [\ion{O}{1}]/H$\alpha$ ratio.}
\label{composite1}
\end{figure}

In Figure \ref{composite1} we again show various optical line ratios for the nuclei and filaments in our sample of 9 clusters, but now with the predicted contribution from the cooling flow removed. We find that the [\ion{N}{2}]/H$\alpha$ and [\ion{S}{2}]/H$\alpha$ ratios are in better agreement with those for typical \ion{H}{2} regions in most systems, with the exception of two systems classified as shock-heated by \cite{mcdonald11b}, based on their H$\alpha$/FUV flux ratios. However, the [\ion{N}{2}]/H$\alpha$ and [\ion{O}{1}]/H$\alpha$ ratios are still too high in most cases. Nevertheless, we feel that this composite model is a useful step forward since it introduces no additional free parameters and tries to account for a source of ionization which we know is present in cool core clusters. The optical line fluxes due to the cooling ICM are tied directly to the X-ray-derived cooling rates and the observed H$\alpha$ luminosities. 

\subsubsection{Composite Models -- Star Formation in a Turbulent Cooling Flow}

\begin{figure}[htb]
\centering
\includegraphics[width=0.95\textwidth]{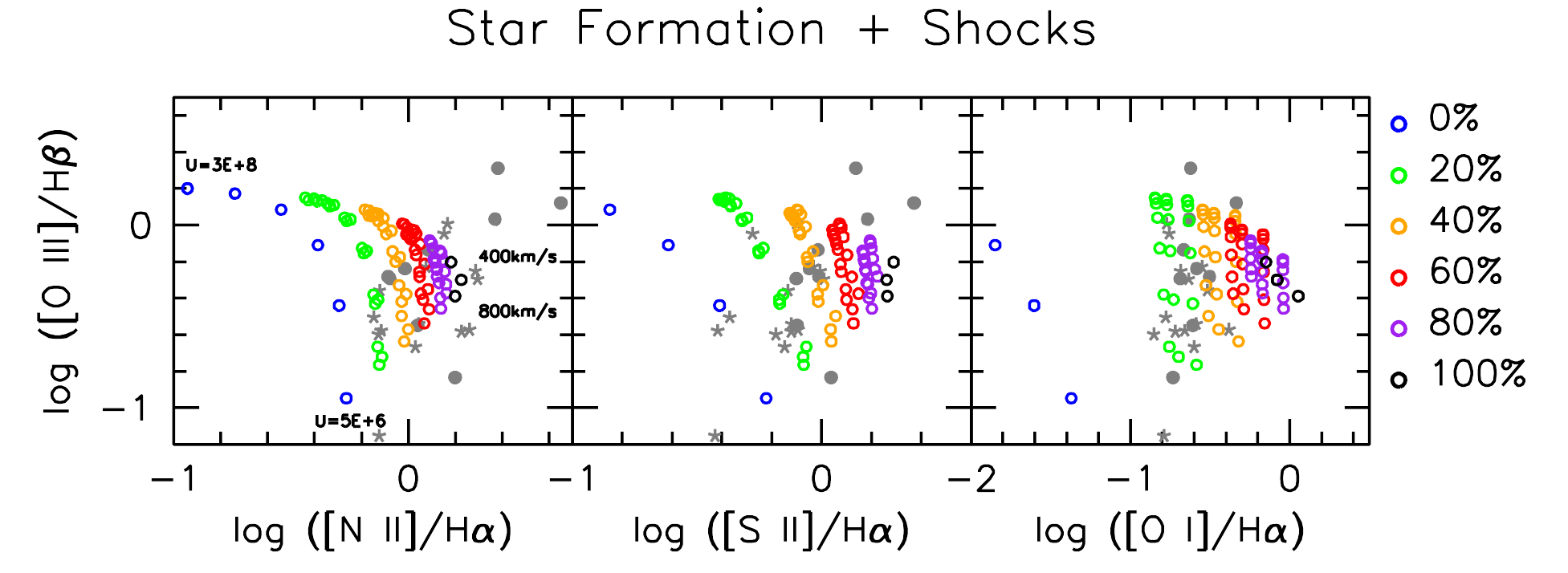}
\caption{Inspired by a similar plot in \cite{rich11}, this figure shows the predicted emission line ratios for a mix of star formation and slow shocks over a range of ionization parameters and shock speed, assuming $Z=Z_{\odot}$. The gray background points are data for filaments (stars) and nuclei (circles) in this sample. The upper panels show the measured flux ratios, while the lower panels show flux ratios with the contribution from a self-ionizing condensing ICM remove (\S4.1.1, Figure \ref{composite1}). These models are able to reproduce the high [\ion{O}{1}]/H$\alpha$ ratios seen in the cool cores of galaxy clusters with a modest ($\sim$0--40\%) amount of shocks.} 
\label{composite_shocks}
\end{figure}

Figure \ref{3x3_vel} shows that the line-emitting gas in cool core clusters generally has a coherent velocity field. The combination of a bulk flow on the order of $v_r \sim 200$ km s$^{-1}$ and line widths of a few times 100 km s$^{-1}$ is reminiscent of the slow shocks combined with star-forming regions seen in large-scale winds \citep{rich10} and in luminous infrared galaxies \citep{rich11}. This composite model, proposed by \cite{farage10} for the optical filaments in Centaurus A, produces slightly-elevated [\ion{N}{2}]/H$\alpha$ and [\ion{S}{2}]/H$\alpha$, and highly-elevated [\ion{O}{1}]/H$\alpha$, just as we observe in the cool cores of galaxy clusters. In Figure \ref{composite_shocks} we show the predicted line ratios from a mix of star formation and radiative shocks from \cite{allen08}. We find that a modest fractional contribution from shocks ($\sim$ 0--40\%) in combination with star formation can produce optical line ratios that match the majority of our data. Further, if we account for self-ionization from the condensing ICM (\S4.1.1), the required contribution from shocks is even lower ($\sim$ 0--20\%). When combined with Figure \ref{sig_lineratios}, this offers compelling support in favor of a composite model including both star formation and shocks.

The observation of two populations in Figure \ref{sig_lineratios}, combined with the above confirmation that a mixture of shocks and star formation is sufficient to produce the observed line ratios, suggests that we can divide our sample into star formation-dominated and shock-dominated subclasses. In Table 2, we summarize the properties of two subsamples: systems which have non-zero UV flux, low [\ion{N}{2}]/H$\alpha$, and a weak or no  correlation between the low ionization line ratios and velocity dispersion (star formation dominated; Abell~0478, Abell~0780, Abell~1795, Abell~2597, Sersic~159-03) and systems for which there is no detectable UV emission, high [\ion{N}{2}]/H$\alpha$ ratios, and a strong correlation between the low ionization line ratios and velocity dispersion (shock dominated; Abell~0496, Abell~1644, Abell~2052, NGC~4325). This table clearly demonstrates the differences between the star formation and shock dominated systems, in terms of their optical line ratios, optical line widths, UV flux and morphology, and nuclear line ratios. We caution, however, that the separation seen in Figure \ref{sig_lineratios} is likely a selection effect and we would expect to see a continuum of systems with shock fractions varying from 0--100\% in a larger, more complete sample.

This scenario offers a straightforward explanation for, it seems, all of the observed properties of optical filaments in cool core clusters. The combination of shocks and star formation account for the soft X-ray, UV and excess blue emission, mid-IR emission, slightly elevated UV/H$\alpha$ ratios \citep{mcdonald11b}, small linewidths, and elevated [\ion{O}{1}]/H$\alpha$ ratios. Including contributions from the cooling ICM reduces the need for shocks by roughly a factor of two, but is not necessary to match the data. Assuming that the \ion{H}{2} regions are flowing along the filaments, they should be experiencing weak shocks due to interactions with the relatively stationary ICM.

\begin{table*}[htb]
\label{properties}
\centering
{\small
\begin{tabular}{l l l}
\hline
 & Star Formation Dominated & Shock Dominated\\
\hline\hline
Systems: & A0478, A0780, A1795, A2597, & A0496, A1644, A2052, N4325 \\
& S159-03 & \\
& & \\
Optical & $< [$\ion{N}{2}$]$/H$\alpha >_{nuc}$ = 0.99 $\pm$ 0.17 & $< [$\ion{N}{2}$]$/H$\alpha >_{nuc}$ = 2.70 $\pm$ 1.24\\
Emission-line & $< [$\ion{S}{2}$]$/H$\alpha >_{nuc}$ = 0.89 $\pm$ 0.09 & $< [$\ion{S}{2}$]$/H$\alpha >_{nuc}$ = 1.59 $\pm$ 0.54\\
Properties: & $< [$\ion{O}{3}$]$/H$\beta >_{nuc}$ = 0.25$\pm$ 0.04 & $< [$\ion{O}{3}$]$/H$\beta >_{nuc}$ = 0.28 $\pm$ 0.12\\
& & \\
& $< [$\ion{N}{2}$]$/H$\alpha >_{fil}$ = 0.85 $\pm$ 0.20 & $< [$\ion{N}{2}$]$/H$\alpha >_{fil}$ = 1.76 $\pm$ 0.33\\
& $< [$\ion{S}{2}$]$/H$\alpha >_{fil}$ = 0.61 $\pm$ 0.19 & $< [$\ion{S}{2}$]$/H$\alpha >_{fil}$ = 0.82 $\pm$ 0.21\\
& & \\
 & $[$\ion{N}{2}$]$/H$\alpha$ $\propto \sigma^{0.22}$ & $[$\ion{N}{2}$]$/H$\alpha$ $\propto \sigma^{1.06}$ \\ 
  & $[$\ion{S}{2}$]$/H$\alpha$ $\propto \sigma^{0.74}$ & $[$\ion{S}{2}$]$/H$\alpha$ $\propto \sigma^{1.32}$ \\ 
  & & \\

  & & \\
 UV Properties$^a$: & UV-H$\alpha$ flux/morphology correlation & Weak UV emission \\
  & & \\
  Nuclear Properties$^b$: & Nuclear starburst & LINER\\
  & & \\
  Shock Fraction$^c$: & $\sim$0--40\% & $\sim$40--100\%\\
  \hline
\end{tabular}
\caption{Various properties of the two sub-samples which were defined based on how strongly the low ionization line ratios were correlated with the velocity dispersion (Figure \ref{sig_lineratios}). 
\newline $^a$: \emph{For systems which have HST far-UV imaging from \cite{mcdonald09} and \cite{mcdonald11b}.}
\newline $^b$: \emph{Based on optical line ratios in Figure \ref{lineratios}.}
\newline $^c$: \emph{Fraction of H$\alpha$ luminosity produced by shocks, based on the composite star formation plus shocks models shown in Figure \ref{composite_shocks}.}
}
%}
}
\end{table*}

\subsection{The Origin of Optical Filaments - Inflow or Outflow?}

The three most likely scenarios for the origin of the cool filaments in cluster cores are radial infall,  buoyant outflow behind radio bubbles, and entrainment in radio jets. These models predict different kinematic signatures in the filaments which should be observable. In the classical picture of a rising bubble, cool material is entrained behind the bubble and rises to larger radius. Directly behind the bubble, the cool material should be moving away from the cluster center, while at small radius the gas which is trailing far behind the bubble cools and falls back into the bottom of the potential well \citep[e.g.,][]{churazov01, reynolds05}. This results in a stretching of the filament, with the inner and outer portions moving in different directions. The line-of-sight velocities of such a filament should show a change of sign along the filament, as is seen in the core of Perseus A \citep{fabian03, hatch06}. While we do not observe a change in direction along any of the extended filaments in our sample, this may simply mean that we are observing the filaments early on, at which time the bulk of the gas will be outflowing, or at a late stage, when all of the cool gas is falling back on the BCG. Indeed, the extended filaments in Abell~2052 and Abell~2597 are coincident with the outer rims of the X-ray cavities, but have nearly-constant line-of-sight velocites.

A second scenario for the extended, cool gas is that it is being uplifted by radio jets. This was recently proposed as the origin of the extended warm gas in Sersic 150-03 \citep{werner11}, but was also investigated much earlier for a sample of radio-selected galaxies by \cite{baum89}. In this scenario, cool gas which has settled in the cluster core is entrained within radio jets during an episode of feedback. Removed from the direct influence of the AGN, this gas is allowed to cool further, resulting in the formation of young stars. The kinematics of the filaments in this case should be relatively smooth, with a constant velocity direction dictated by the radio jet. This is consistent with our observations for the most extended systems, as discussed in \S3.2. This process is likely occurring in Sersic 159-03 \citep{werner11}, and at small radii in Abell~1795 \citep{vanbreugel84, mcdonald09}. 

Gravitational freefall of a cool cloud may be the simplest of these three models, and is likely happening to some extent in most systems. Assuming a particle is in freefall through an atmosphere, the radial velocity should rise to a maximum speed at large radius where the atmosphere density is minimal, decelerate at intermediate radii as it experiences increasing drag, and then accelerate at small radii when gravity dominates once again. The particle should have the same velocity direction at all radii, unlike the buoyant bubble scenario. This characteristic shape is seen in the extended filaments of Abell~0496 and Abell~1795. Of the three aforementioned scenarios, this may be the most difficult to confirm, as there is no obvious secondary signature (i.e. radio emission, X-ray cavities). 

Without additional information, line-of-sight kinematics are insufficient to differentiate between the three aforementioned models. Seeking to identify whether optically-emitting gas is in the foreground or background of the BCG is critical to interpreting these velocities. Additionally, further work identifying the small-scale (arcsecond) orientation of radio jets in cluster cores and low surface brightness x-ray cavities would provide additional evidence in favor or against the two outflow models. 

%================================================================%
%============== SUMMARY ========================================%
%================================================================%

\section{Summary}
Utilizing a combination of narrow-band H$\alpha$ imaging and carefully-positioned long-slit spectroscopy we have obtained optical spectra of H$\alpha$ filaments in the cool cores of 9 galaxy clusters. These spectra provide a wealth of information about the warm, ionized gas in cool core clusters, allowing us to provide the following new constraints on their origin and ionization mechanisms:

\begin{itemize}
\item We find evidence for only small amounts of reddening in the extended, line-emitting filaments, with the majority of filaments having $E(B-V)<0.2$.
\item We confirm that the optical line ratios [\ion{N}{2}] $\lambda$6583/H$\alpha$, [\ion{O}{1}] $\lambda$6300/H$\alpha$, [\ion{S}{2}] $\lambda\lambda$6716, 6731/H$\alpha$ are slightly elevated above the expectation for star-forming regions. We find that filaments identified as star-forming by their far-UV emission consistently have the lowest [\ion{N}{2}]/H$\alpha$ ratios.
\item For the complete sample there is a weak correlation between the low-ionization line ratios and line widths. This correlation is stronger for systems which appear to have a larger contribution from shocks than from star formation, based on their optical line ratios and far-UV/H$\alpha$ flux ratios. 
\item The optical line ratios are able to rule out collisional ionization by cosmic rays, thermal conduction, and photoionization by both the ICM X-rays and AGN as strong contributors to the ionization of the warm gas in both nuclei and filaments.
\item A composite model of a self-ionizing cooling plasma and star formation is able to better reproduce these line ratios than star formation alone, but still under-predicts the low-ionization ratios in most systems.
\item The data are fully consistent with a combination of slow shocks and star formation, as seen in superwinds and merger remnants. This scenario explains the kinematics, line ratios, and multi-wavelength properties of the filaments and nuclei. We find typical shock fractions of 0--40\% in systems with signatures of star formation (i.e. far-UV emission), and 40--100\% in systems with no observational evidence for star formation. Including self-ionization from the cooling plasma to this model reduces the necessary contribution from shocks by roughly a factor of 2 while still matching the data.
\item The nuclei of Abell~0496, Abell~1644, and Abell~2052 are identified as LINERs. The remaining 6 nuclei in the sample share similar properties to the star-forming filaments and likely have very little contribution from an AGN.
\item In the central ($r<10$ kpc) regions of several cool core clusters, we see kinematic signatures of rotating, H$\alpha$ structures, consistent with earlier work \citep[e.g.,][]{baum92}. 
\item The line-of-sight velocities of the most extended filaments peak at $\sim$300 km s$^{-1}$. We do not observe a change in direction along any filaments. We provide evidence drawn from the literature for entrained gas in radio jets (Sersic~159-03, Abell~1795), cool gas uplifted by buoyant bubbles (Abell~2052, Abell~2597), and gravitational freefall (Abell~0496, Abell~1795).
\item The line profiles of the ionized filaments have widths on the order of FWHM $\sim 200$ km s$^{-1}$. This is consistent with velocity broadening due to slow shocks. In general, filaments that are thin and extended have narrower emission lines than those that appear disturbed.
\item The line widths in the filaments decrease with distance from the BCG center. If the cool filaments have properties which reflect the ICM environment, this provides an estimate of the radial velocity dispersion profile in the ICM. These velocities widths are consistent with those measured for the ICM using X-ray grating spectrometers \citep[e.g., ][]{sanders11} and are much smaller than the stellar velocity dispersions in the BCG.
\end{itemize}

These new data, coupled with our previous work \citep{mcdonald10,mcdonald11a,mcdonald11b}, support a scenario where some fraction of the cooling ICM in the cluster core has been allowed to collapse and form stars. While in bulk motion (either infall or outflow), the cool gas is experiencing shocks, producing the observed line widths of $\sim$ 200 km s$^{-1}$, which are significantly higher than those measured in typical \ion{H}{2} regions. Further, the cooling ICM may contribute a non-negligible fraction of self-ionization as it cools through EUV and UV transitions. In filaments which appear to be star-forming based on their FUV and IR emission, the combination of shocks and self-ionization yield 0--40\% of the H$\alpha$ luminosity in the filaments, with photoionization by young stars providing the additional 60--100\%. Systems without any accompanying FUV emission appear to be fully consistent with ionization by radiative shocks. While satisfactory, this scenario is necessarily incomplete. A complete modeling of the multiphase plasma using all available data from X-ray through radio, rather than focusing on a single energy regime, should  allow for a more robust solution to this long-standing problem, while combining the warm gas kinematics with deep X-ray and radio studies may shed new light on the motions of this gas.

\section*{Acknowledgements}
Support for this work was provided to M.M. by NASA through SAO Award Number 2834-MIT-SAO-4018, which is issued by the Chandra X-ray Observatory on behalf of NASA under contract NAS8-03060, and to both MM and S.V.\ by NSF through contracts AST 0606932 and 1009583, and by NASA through contract HST GO-1198001A. We thank R.\ Mushotzky and A.\ Edge for useful discussions.  We also thank the technical staff at Las Campanas for their support during the ground-based observations.

\renewcommand{\thetable}{A-\arabic{table}}
\setcounter{table}{0}

\appendix
\begin{landscape}
\section*{Appendix A}
%\begin{appendix}

We provide here the results of our spectroscopic analysis for each slit position. For each spatial element, we provide the radius from the BCG nucleus, the line-of-sight velocity relative to the BCG, the FWHM of the H$\alpha$ emission line and fluxes for the H$\alpha$, H$\beta$, [\ion{O}{3}], [\ion{O}{1}], [\ion{N}{2}], and [\ion{S}{2}] lines. These lines are uncorrected for any reddening. Spatial bins which have no measurable optical emission are not shown, for brevity.
%\clearpage
%\addtolength{\oddsidemargin}{-.7in}
%\addtolength{\evensidemargin}{-.7in}
%\addtolength{\textwidth}{1.4in}
\begin{center}
{\scriptsize
\begin{longtable}{|c c c c | r r | r r r r r r |}
\caption{Kinematics and fluxes (uncorrected for reddening) derived from optical spectra for each spatial element along both slits for each galaxy cluster.}
\label{datatable} \\

\hline \hline

Name & Slit & Spatial & Radius & \multicolumn{1}{c}{$v_{rad}$} & \multicolumn{1}{c|}{FWHM} & \multicolumn{1}{c}{F$_{H\alpha}$} & \multicolumn{1}{c}{F$_{H\beta}$} & \multicolumn{1}{c}{F$_{[O III]}$} & \multicolumn{1}{c}{F$_{[O I]}$} & \multicolumn{1}{c}{F$_{[N II]}$} & \multicolumn{1}{c|}{F$_{[S II]}$}\\
 & No. & Bin & [kpc] & \multicolumn{2}{|c|}{[km s$^{-1}$]} & \multicolumn{6}{c|}{[10$^{-14} erg s^{-1}$]}\\
\hline

\endfirsthead

\multicolumn{12}{c}%
{{\bfseries \tablename\ \thetable{} -- continued from previous page}} \\

\hline \hline
Name & Slit & Spatial & Radius & \multicolumn{1}{c}{$v_{rad}$} & \multicolumn{1}{c|}{FWHM} & \multicolumn{1}{c}{F$_{H\alpha}$} & \multicolumn{1}{c}{F$_{H\beta}$} & \multicolumn{1}{c}{F$_{[O III]}$} & \multicolumn{1}{c}{F$_{[O I]}$} & \multicolumn{1}{c}{F$_{[N II]}$} & \multicolumn{1}{c|}{F$_{[S II]}$}\\
 & No. & Bin & [kpc] & \multicolumn{2}{|c|}{[km s$^{-1}$]} & \multicolumn{6}{c|}{[10$^{-14} erg s^{-1}$]}\\
\hline
\endhead

\hline \multicolumn{12}{| l |}{{Continued on next page}} \\ \hline
\endfoot

\hline \hline
\endlastfoot

A1795 & 1 & 2 & 45.1 & 144(10) & 46(5) & 0.009(0.001) & 0.007(0.001) & 0.000(0.001) & 0.000(0.001) & 0.005(0.001) & 0.003(0.001) \\ 
  &  &  3 & 41.3 & 142(10) & 40(5) & 0.035(0.001) & 0.008(0.001) & 0.003(0.001) & 0.007(0.001) & 0.023(0.001) & 0.011(0.001) \\ 
  &  &  4 & 37.6 & 112(10) & 42(5) & 0.023(0.001) & 0.007(0.001) & 0.002(0.001) & 0.005(0.001) & 0.018(0.001) & 0.010(0.001) \\ 
  &  &  5 & 33.8 & 89(10) & 46(5) & 0.026(0.001) & 0.007(0.001) & 0.003(0.001) & 0.007(0.001) & 0.023(0.001) & 0.013(0.001) \\ 
  &  &  6 & 30.0 & 87(10) & 43(5) & 0.091(0.001) & 0.027(0.001) & 0.008(0.001) & 0.015(0.001) & 0.063(0.001) & 0.039(0.001) \\ 
  &  &  7 & 26.2 & 76(10) & 46(5) & 0.068(0.001) & 0.013(0.001) & 0.005(0.001) & 0.005(0.001) & 0.043(0.001) & 0.027(0.001) \\ 
  &  &  10 & 14.9 & 40(11) & 47(6) & 0.005(0.001) & 0.003(0.001) & 0.000(0.001) & 0.001(0.001) & 0.005(0.001) & 0.003(0.001) \\ 
  &  &  11 & 11.1 & 34(11) & 55(5) & 0.006(0.001) & 0.002(0.001) & 0.000(0.001) & 0.002(0.001) & 0.005(0.001) & 0.002(0.001) \\ 
  &  &  12 & 7.3 & 29(10) & 66(4) & 0.016(0.001) & 0.005(0.001) & 0.002(0.001) & 0.001(0.001) & 0.013(0.001) & 0.008(0.001) \\ 
  &  &  13 & 3.6 & 152(10) & 165(2) & 0.222(0.001) & 0.086(0.001) & 0.051(0.001) & 0.058(0.001) & 0.210(0.001) & 0.194(0.001) \\ 
  &  &  14 & 0.2 & 243(10) & 234(1) & 0.599(0.002) & 0.102(0.003) & 0.067(0.001) & 0.162(0.001) & 0.620(0.002) & 0.590(0.002) \\ 
  &  &  15 & 4.0 & -179(10) & 125(2) & 0.285(0.001) & 0.079(0.002) & 0.036(0.001) & 0.068(0.001) & 0.239(0.001) & 0.204(0.001) \\ 
  &  &  16 & 7.8 & -218(10) & 152(2) & 0.044(0.001) & 0.004(0.001) & 0.002(0.001) & 0.011(0.001) & 0.040(0.001) & 0.029(0.001) \\ 
  & 2 & 1 & 56.1 & 206(10) & 58(4) & 0.068(0.001) & 0.019(0.001) & 0.003(0.001) & 0.012(0.001) & 0.037(0.001) & 0.020(0.001) \\ 
  &  &  2 & 52.3 & 176(10) & 42(5) & 0.034(0.001) & 0.011(0.001) & 0.001(0.001) & 0.006(0.001) & 0.024(0.001) & 0.012(0.001) \\ 
  &  &  3 & 48.6 & 169(10) & 41(5) & 0.028(0.001) & 0.008(0.001) & 0.000(0.001) & 0.005(0.001) & 0.020(0.001) & 0.009(0.001) \\ 
  &  &  4 & 44.8 & 160(10) & 40(5) & 0.013(0.001) & 0.002(0.001) & 0.001(0.001) & 0.002(0.001) & 0.009(0.001) & 0.005(0.001) \\ 
  &  &  5 & 41.1 & 149(10) & 40(5) & 0.008(0.001) & 0.003(0.001) & 0.000(0.001) & 0.001(0.001) & 0.007(0.001) & 0.002(0.001) \\ 
  &  &  6 & 37.3 & 139(10) & 48(5) & 0.008(0.001) & 0.002(0.001) & 0.001(0.001) & 0.001(0.001) & 0.008(0.001) & 0.005(0.001) \\ 
  &  &  7 & 33.5 & 138(10) & 41(5) & 0.015(0.001) & 0.004(0.001) & 0.001(0.001) & 0.003(0.001) & 0.012(0.001) & 0.006(0.001) \\ 
  &  &  8 & 29.8 & 165(10) & 64(4) & 0.018(0.001) & 0.005(0.001) & 0.002(0.001) & 0.004(0.001) & 0.014(0.001) & 0.006(0.001) \\ 
  &  &  9 & 26.0 & 184(10) & 56(4) & 0.012(0.001) & 0.003(0.001) & 0.001(0.001) & 0.002(0.001) & 0.010(0.001) & 0.003(0.001) \\ 
  &  &  10 & 22.3 & 231(10) & 63(4) & 0.007(0.001) & 0.003(0.001) & 0.002(0.001) & 0.002(0.001) & 0.008(0.001) & 0.003(0.001) \\ 
  &  &  11 & 18.6 & -98(10) & 74(4) & 0.011(0.001) & 0.004(0.001) & 0.001(0.001) & 0.004(0.001) & 0.010(0.001) & 0.005(0.001) \\ 
  &  &  11 & 18.6 & 155(262) & 50(5) & 0.012(0.001) & 0.004(0.001) & 0.000(0.001) & 0.003(0.001) & 0.011(0.001) & 0.005(0.001) \\ 
  &  &  12 & 14.9 & -108(12) & 109(6) & 0.014(0.001) & 0.004(0.001) & 0.002(0.001) & 0.002(0.001) & 0.012(0.001) & 0.006(0.001) \\ 
  &  &  12 & 14.9 & 144(257) & 60(5) & 0.011(0.001) & 0.003(0.001) & 0.001(0.001) & 0.002(0.001) & 0.010(0.001) & 0.006(0.001) \\ 
  &  &  13 & 11.4 & 140(10) & 75(3) & 0.031(0.001) & 0.009(0.001) & 0.003(0.001) & 0.008(0.001) & 0.026(0.001) & 0.016(0.001) \\ 
A0780 & 1 & 1 & 3.4 & -148(11) & 146(6) & 0.017(0.001) & 0.018(0.001) & 0.001(0.001) & 0.004(0.001) & 0.012(0.001) & 0.010(0.001) \\ 
  &  &  2 & 2.3 & -193(10) & 109(2) & 0.160(0.001) & 0.067(0.001) & 0.016(0.001) & 0.021(0.001) & 0.111(0.001) & 0.099(0.001) \\ 
  &  &  3 & 1.2 & -136(10) & 108(2) & 0.238(0.001) & 0.049(0.001) & 0.019(0.001) & 0.043(0.001) & 0.181(0.001) & 0.173(0.001) \\ 
  &  &  4 & 0.2 & 7(10) & 145(2) & 0.190(0.001) & 0.055(0.002) & 0.031(0.001) & 0.044(0.001) & 0.165(0.001) & 0.156(0.001) \\ 
  &  &  5 & 1.0 & 204(10) & 174(2) & 0.193(0.001) & 0.047(0.001) & 0.025(0.001) & 0.039(0.001) & 0.172(0.001) & 0.163(0.001) \\ 
  &  &  6 & 2.0 & 397(10) & 130(2) & 0.135(0.001) & 0.024(0.001) & 0.009(0.001) & 0.018(0.001) & 0.103(0.001) & 0.094(0.001) \\ 
  &  &  7 & 3.1 & 433(10) & 119(4) & 0.016(0.001) & 0.003(0.001) & 0.000(0.001) & 0.002(0.001) & 0.015(0.001) & 0.012(0.001) \\ 
  &  &  8 & 4.2 & 75(10) & 52(5) & 0.004(0.001) & 0.001(0.001) & 0.001(0.001) & 0.000(0.001) & 0.005(0.001) & 0.002(0.001) \\ 
  & 2 & 3 & 3.7 & -64(10) & 64(3) & 0.013(0.001) & 0.005(0.001) & 6.025(0.001) & 0.003(0.001) & 0.013(0.001) & 0.010(0.001) \\ 
  &  &  4 & 3.8 & -74(10) & 87(3) & 0.015(0.001) & 0.004(0.001) & 0.002(0.001) & 0.003(0.001) & 0.016(0.001) & 0.012(0.001) \\ 
  &  &  5 & 4.2 & -47(10) & 105(3) & 0.012(0.001) & 0.003(0.001) & 0.005(0.001) & 0.004(0.001) & 0.015(0.001) & 0.011(0.001) \\ 
  &  &  6 & 4.8 & -46(10) & 89(3) & 0.015(0.001) & 0.006(0.001) & 0.005(0.001) & 0.004(0.001) & 0.020(0.001) & 0.017(0.001) \\ 
  &  &  7 & 5.6 & -85(10) & 94(3) & 0.013(0.001) & 0.005(0.001) & 0.000(0.001) & 0.003(0.001) & 0.018(0.001) & 0.013(0.001) \\ 
A1644 & 1 & 1 & 3.7 & -96(10) & 106(3) & 0.019(0.001) & 0.013(0.002) & 0.013(0.001) & 0.007(0.001) & 0.072(0.001) & 0.033(0.001) \\ 
  &  &  2 & 0.9 & 38(10) & 180(2) & 0.136(0.001) & 0.074(0.007) & 0.103(0.002) & 0.064(0.001) & 0.628(0.002) & 0.335(0.002) \\ 
  &  &  4 & 4.6 & 105(10) & 72(4) & 0.016(0.001) & 0.009(0.001) & 8.895(0.001) & 0.002(0.001) & 0.038(0.001) & 0.018(0.001) \\ 
  & 2 & 1 & 6.3 & -209(10) & 60(5) & 0.014(0.001) & 0.011(0.001) & 0.000(0.001) & 0.001(0.001) & 0.009(0.001) & 0.004(0.001) \\ 
  &  &  2 & 6.1 & -184(10) & 74(4) & 0.009(0.001) & 0.005(0.001) & 0.003(0.001) & 0.002(0.001) & 0.018(0.001) & 0.009(0.001) \\ 
  &  &  3 & 7.1 & -186(10) & 74(3) & 0.011(0.001) & 0.009(0.001) & 0.004(0.001) & 0.001(0.001) & 0.027(0.001) & 0.012(0.001) \\ 
  &  &  4 & 8.8 & -184(10) & 63(4) & 0.014(0.001) & 0.005(0.001) & 0.000(0.001) & 0.004(0.001) & 0.028(0.001) & 0.011(0.001) \\ 
A2052 & 1 & 1 & 5.9 & 179(11) & 99(5) & 0.012(0.001) & 0.003(0.001) & 0.003(0.001) & 0.007(0.001) & 0.026(0.001) & 0.012(0.001) \\ 
  &  &  2 & 3.8 & 149(11) & 105(5) & 0.016(0.001) & 0.000(0.001) & 0.000(0.001) & 0.003(0.001) & 0.025(0.001) & 0.009(0.001) \\ 
  &  &  3 & 1.7 & 76(10) & 111(3) & 0.029(0.001) & 0.016(0.002) & 0.014(0.001) & 0.007(0.001) & 0.075(0.001) & 0.039(0.001) \\ 
  &  &  4 & 0.5 & 16(10) & 112(2) & 0.152(0.001) & 0.032(0.002) & 0.082(0.001) & 0.039(0.001) & 0.371(0.002) & 0.211(0.002) \\ 
  &  &  5 & 2.4 & -57(10) & 101(4) & 0.020(0.001) & 0.000(0.001) & 0.002(0.001) & 0.001(0.001) & 0.040(0.001) & 0.026(0.001) \\ 
  &  &  6 & 4.5 & 50(10) & 82(3) & 0.019(0.001) & 0.010(0.001) & 0.004(0.001) & 0.000(0.001) & 0.037(0.001) & 0.014(0.001) \\ 
  & 2 & 2 & 13.8 & 253(10) & 130(2) & 0.028(0.001) & 0.004(0.001) & 0.005(0.001) & 0.005(0.001) & 0.034(0.001) & 0.012(0.001) \\ 
  &  &  3 & 12.6 & 253(10) & 83(3) & 0.034(0.001) & 0.005(0.001) & 0.005(0.001) & 0.005(0.001) & 0.050(0.001) & 0.019(0.001) \\ 
  &  &  4 & 11.6 & 200(10) & 83(3) & 0.020(0.001) & 0.005(0.001) & 0.002(0.001) & 0.003(0.001) & 0.033(0.001) & 0.011(0.001) \\ 
S15903 & 1 & 1 & 16.6 & 251(10) & 81(3) & 0.193(0.002) & 0.058(0.002) & 0.005(0.002) & 0.031(0.001) & 0.136(0.001) & 0.067(0.002) \\ 
  &  &  2 & 12.0 & 178(10) & 97(2) & 0.135(0.002) & 0.041(0.003) & 0.004(0.002) & 0.018(0.001) & 0.105(0.001) & 0.059(0.002) \\ 
  &  &  3 & 7.5 & 143(11) & 110(4) & 0.045(0.001) & 0.014(0.003) & 0.004(0.002) & 0.012(0.001) & 0.042(0.001) & 0.016(0.002) \\ 
  &  &  4 & 2.8 & 158(10) & 115(3) & 0.094(0.002) & 0.012(0.003) & 0.011(0.002) & 0.020(0.001) & 0.123(0.002) & 0.080(0.002) \\ 
  &  &  5 & 1.7 & 334(10) & 196(1) & 0.501(0.003) & 0.092(0.004) & 0.058(0.003) & 0.103(0.002) & 0.586(0.003) & 0.474(0.004) \\ 
  & 2 & 1 & 6.0 & 389(10) & 138(3) & 0.068(0.001) & 0.030(0.003) & 0.000(0.002) & 0.019(0.001) & 0.079(0.002) & 0.043(0.002) \\ 
  &  &  2 & 5.2 & 412(10) & 103(2) & 0.212(0.001) & 0.059(0.002) & 0.014(0.002) & 0.040(0.001) & 0.185(0.001) & 0.118(0.002) \\ 
  &  &  3 & 6.6 & 368(10) & 96(2) & 0.207(0.001) & 0.052(0.003) & 0.003(0.001) & 0.041(0.001) & 0.167(0.001) & 0.091(0.001) \\ 
  &  &  4 & 9.1 & 354(10) & 106(3) & 0.087(0.001) & 0.026(0.002) & 0.005(0.002) & 0.016(0.001) & 0.073(0.001) & 0.037(0.001) \\ 
  &  &  5 & 12.2 & 264(10) & 70(4) & 0.036(0.001) & 0.013(0.002) & 0.000(0.001) & 0.007(0.001) & 0.032(0.001) & 0.015(0.001) \\ 
  &  &  6 & 15.4 & 261(10) & 64(3) & 0.073(0.001) & 0.018(0.002) & 0.000(0.001) & 0.012(0.001) & 0.053(0.001) & 0.033(0.001) \\ 
  &  &  7 & 18.7 & 292(10) & 52(4) & 0.074(0.001) & 0.015(0.002) & 0.000(0.001) & 0.011(0.001) & 0.046(0.001) & 0.029(0.001) \\ 
  &  &  8 & 22.1 & 300(10) & 48(5) & 0.035(0.001) & 0.019(0.002) & 0.000(0.001) & 0.008(0.001) & 0.027(0.001) & 0.009(0.001) \\ 
  &  &  9 & 25.4 & 298(10) & 56(4) & 0.040(0.001) & 0.010(0.003) & 0.000(0.001) & 0.006(0.001) & 0.023(0.001) & 0.018(0.001) \\ 
A0496 & 1 & 2 & 3.8 & 29(10) & 89(4) & 0.009(0.001) & 0.002(0.001) & 0.000(0.001) & 0.000(0.001) & 0.013(0.001) & 0.006(0.001) \\ 
  &  &  3 & 2.9 & 38(10) & 123(3) & 0.015(0.001) & 0.006(0.001) & 0.002(0.001) & 0.002(0.001) & 0.026(0.001) & 0.014(0.001) \\ 
  &  &  4 & 2.0 & 135(10) & 133(3) & 0.043(0.001) & 0.005(0.001) & 0.003(0.001) & 0.008(0.001) & 0.091(0.001) & 0.044(0.001) \\ 
  &  &  5 & 1.1 & 175(10) & 107(2) & 0.130(0.002) & 0.018(0.001) & 0.014(0.001) & 0.029(0.001) & 0.270(0.002) & 0.150(0.001) \\ 
  &  &  6 & 0.2 & 178(10) & 155(2) & 0.254(0.002) & 0.049(0.005) & 0.046(0.001) & 0.065(0.001) & 0.587(0.003) & 0.384(0.002) \\ 
  &  &  7 & 0.7 & 149(10) & 158(2) & 0.127(0.003) & 0.028(0.001) & 0.037(0.001) & 0.023(0.002) & 0.307(0.003) & 0.200(0.002) \\ 
  &  &  8 & 1.6 & 55(10) & 146(2) & 0.063(0.002) & 0.010(0.001) & 0.008(0.001) & 0.011(0.001) & 0.116(0.001) & 0.065(0.001) \\ 
  &  &  9 & 2.5 & 73(10) & 93(2) & 0.044(0.001) & 0.008(0.001) & 0.005(0.001) & 0.010(0.001) & 0.074(0.001) & 0.036(0.001) \\ 
  &  &  10 & 3.4 & 79(10) & 101(2) & 0.026(0.001) & 0.009(0.001) & 0.003(0.001) & 0.007(0.001) & 0.045(0.001) & 0.022(0.001) \\ 
  &  &  11 & 4.3 & 182(10) & 123(4) & 0.008(0.001) & 0.003(0.001) & 0.000(0.001) & 0.001(0.001) & 0.016(0.001) & 0.007(0.001) \\ 
  &  &  12 & 5.2 & 258(10) & 93(4) & 0.005(0.001) & 0.002(0.001) & 0.000(0.001) & 0.000(0.001) & 0.014(0.001) & 0.004(0.001) \\ 
  &  &  13 & 6.1 & 221(10) & 96(4) & 0.007(0.001) & 0.000(0.001) & 0.000(0.001) & 0.000(0.001) & 0.013(0.001) & 0.004(0.001) \\ 
  &  &  14 & 7.0 & 179(10) & 76(4) & 0.003(0.001) & 0.000(0.001) & 0.000(0.001) & 0.001(0.001) & 0.008(0.001) & 0.000(0.001) \\ 
  &  &  15 & 7.9 & 146(12) & 44(8) & 0.002(0.001) & 0.000(0.001) & 0.000(0.001) & 0.000(0.001) & 0.002(0.001) & -1.21(0.001) \\ 
  & 2 & 3 & 3.4 & -12(11) & 85(5) & 0.002(0.001) & 0.000(0.001) & 0.000(0.001) & 0.001(0.001) & 0.011(0.001) & 0.006(0.001) \\ 
  &  &  4 & 2.7 & 27(10) & 132(3) & 0.016(0.001) & 0.001(0.001) & 0.002(0.001) & 0.003(0.001) & 0.044(0.001) & 0.025(0.001) \\ 
  &  &  5 & 2.0 & 105(10) & 106(2) & 0.074(0.001) & 0.020(0.001) & 0.011(0.001) & 0.017(0.001) & 0.156(0.001) & 0.084(0.001) \\ 
  &  &  6 & 1.6 & 142(10) & 89(2) & 0.160(0.001) & 0.034(0.001) & 0.019(0.001) & 0.036(0.001) & 0.311(0.001) & 0.155(0.001) \\ 
  &  &  7 & 1.7 & 173(10) & 118(2) & 0.138(0.001) & 0.026(0.001) & 0.014(0.001) & 0.031(0.001) & 0.270(0.001) & 0.143(0.001) \\ 
  &  &  8 & 2.1 & 239(10) & 90(3) & 0.071(0.001) & 0.014(0.001) & 0.005(0.001) & 0.017(0.001) & 0.143(0.001) & 0.074(0.001) \\ 
  &  &  9 & 2.8 & 238(10) & 66(3) & 0.057(0.001) & 0.011(0.001) & 0.004(0.001) & 0.011(0.001) & 0.104(0.001) & 0.049(0.001) \\ 
  &  &  10 & 3.6 & 228(10) & 58(4) & 0.038(0.001) & 0.006(0.001) & 0.001(0.001) & 0.010(0.001) & 0.064(0.001) & 0.031(0.001) \\ 
  &  &  11 & 4.4 & 202(10) & 73(4) & 0.009(0.001) & 0.001(0.001) & 0.000(0.001) & 0.002(0.001) & 0.019(0.001) & 0.010(0.001) \\ 
A0478 & 1 & 2 & 6.9 & -161(10) & 138(3) & 0.016(0.001) & 0.003(0.001) & 0.000(0.001) & 0.004(0.001) & 0.017(0.001) & 0.011(0.001) \\ 
  &  &  3 & 4.6 & -68(10) & 142(2) & 0.064(0.001) & 0.010(0.001) & 0.002(0.001) & 0.017(0.001) & 0.074(0.001) & 0.052(0.001) \\ 
  &  &  4 & 2.2 & -12(10) & 140(2) & 0.120(0.001) & 0.016(0.001) & 0.007(0.001) & 0.031(0.001) & 0.137(0.001) & 0.104(0.001) \\ 
  &  &  5 & 0.7 & 26(10) & 159(1) & 0.233(0.001) & 0.033(0.001) & 0.010(0.001) & 0.056(0.001) & 0.257(0.001) & 0.198(0.001) \\ 
  &  &  6 & 2.7 & 103(10) & 134(2) & 0.167(0.001) & 0.022(0.001) & 0.007(0.001) & 0.041(0.001) & 0.185(0.001) & 0.131(0.001) \\ 
  &  &  7 & 5.0 & 121(10) & 126(2) & 0.089(0.001) & 0.013(0.001) & 0.002(0.001) & 0.022(0.001) & 0.100(0.001) & 0.061(0.001) \\ 
  &  &  8 & 7.3 & 75(10) & 109(2) & 0.058(0.001) & 0.008(0.001) & 0.001(0.001) & 0.014(0.001) & 0.059(0.001) & 0.037(0.001) \\ 
  &  &  9 & 9.7 & 71(10) & 113(3) & 0.019(0.001) & 0.002(0.001) & 0.000(0.001) & 0.003(0.001) & 0.019(0.001) & 0.011(0.001) \\ 
  & 2 & 2 & 7.1 & -85(12) & 207(7) & 0.011(0.001) & 0.002(0.001) & 0.000(0.001) & 0.002(0.001) & 0.009(0.001) & 0.005(0.001) \\ 
  &  &  3 & 5.0 & -45(10) & 186(3) & 0.022(0.001) & 0.004(0.001) & 0.000(0.001) & 0.006(0.001) & 0.022(0.001) & 0.013(0.001) \\ 
  &  &  4 & 3.3 & 20(10) & 146(2) & 0.078(0.001) & 0.012(0.001) & 0.003(0.001) & 0.022(0.001) & 0.088(0.001) & 0.059(0.001) \\ 
  &  &  5 & 2.8 & 17(10) & 139(2) & 0.211(0.001) & 0.029(0.001) & 0.008(0.001) & 0.052(0.001) & 0.228(0.001) & 0.165(0.001) \\ 
  &  &  6 & 4.0 & 41(10) & 126(2) & 0.182(0.001) & 0.026(0.001) & 0.005(0.001) & 0.045(0.001) & 0.200(0.001) & 0.138(0.001) \\ 
  &  &  7 & 5.9 & 69(10) & 155(1) & 0.155(0.001) & 0.021(0.001) & 0.005(0.001) & 0.037(0.001) & 0.167(0.001) & 0.114(0.001) \\ 
  &  &  8 & 8.1 & 111(10) & 143(2) & 0.058(0.001) & 0.009(0.001) & 0.002(0.001) & 0.013(0.001) & 0.063(0.001) & 0.040(0.001) \\ 
  &  &  9 & 10.3 & 157(11) & 165(4) & 0.013(0.001) & 0.003(0.001) & 0.000(0.001) & 0.003(0.001) & 0.013(0.001) & 0.008(0.001) \\ 
A2597 & 1 & 2 & 18.4 & 99(11) & 75(6) & 0.010(0.001) & 0.005(0.001) & 0.000(0.001) & 0.002(0.001) & 0.007(0.001) & 0.002(0.001) \\ 
  &  &  3 & 15.1 & 125(11) & 77(6) & 0.015(0.001) & 0.004(0.002) & 0.000(0.001) & 0.002(0.001) & 0.005(0.001) & 0.008(0.001) \\ 
  &  &  4 & 11.9 & 85(11) & 94(5) & 0.014(0.001) & 0.001(0.001) & 0.000(0.001) & 0.004(0.001) & 0.008(0.001) & 0.006(0.001) \\ 
  &  &  5 & 8.5 & 54(11) & 93(5) & 0.014(0.001) & 0.009(0.002) & 0.000(0.001) & 0.005(0.001) & 0.010(0.001) & 0.009(0.001) \\ 
  &  &  6 & 5.3 & 217(19) & 267(13) & 0.026(0.001) & 0.023(0.170) & 0.017(0.002) & 0.009(0.001) & 0.012(0.001) & 0.029(0.002) \\ 
  &  &  7 & 2.0 & 41(10) & 289(1) & 0.886(0.003) & 0.242(0.005) & 0.109(0.003) & 0.304(0.002) & 0.766(0.002) & 0.925(0.003) \\ 
  &  &  8 & 1.3 & 93(10) & 259(1) & 1.113(0.002) & 0.255(0.004) & 0.149(0.003) & 0.323(0.001) & 0.872(0.002) & 1.022(0.003) \\ 
  &  &  9 & 4.6 & 60(10) & 192(1) & 0.347(0.001) & 0.088(0.004) & 0.046(0.002) & 0.107(0.001) & 0.277(0.001) & 0.281(0.001) \\ 
  &  &  10 & 7.9 & 154(10) & 123(3) & 0.055(0.001) & 0.017(0.003) & 0.006(0.002) & 0.017(0.001) & 0.037(0.001) & 0.034(0.001) \\ 
  &  &  11 & 11.2 & 212(10) & 102(3) & 0.038(0.001) & 0.014(0.003) & 0.003(0.001) & 0.011(0.001) & 0.025(0.001) & 0.021(0.001) \\ 
  &  &  12 & 14.5 & 268(11) & 65(6) & 0.011(0.001) & 0.006(0.001) & 0.002(0.001) & 0.003(0.001) & 0.004(0.001) & 0.004(0.001) \\ 
  &  &  13 & 17.8 & 257(10) & 44(6) & 0.009(0.001) & 0.002(0.001) & 0.002(0.001) & 0.002(0.001) & 0.004(0.001) & 0.001(0.001) \\ 
  &  &  14 & 21.1 & 239(11) & 44(6) & 0.008(0.001) & 0.004(0.001) & 0.000(0.001) & 0.001(0.001) & 0.003(0.001) & 0.002(0.001) \\ 
N4325 & 1 & 2 & 2.9 & -168(28) & 96(21) & 0.005(0.001) & 0.000(0.001) & 0.007(0.001) & 0.000(0.001) & 0.012(0.003) & 0.012(0.001) \\ 
  &  &  2 & 2.9 & -25(117) & 43(10) & 0.004(0.001) & 0.003(0.001) & 0.000(0.001) & 0.000(0.001) & 0.005(0.002) & 0.003(0.001) \\ 
  &  &  3 & 1.8 & -49(10) & 62(4) & 0.016(0.001) & 0.005(0.001) & 0.000(0.001) & 0.001(0.001) & 0.019(0.001) & 0.011(0.001) \\ 
  &  &  4 & 0.7 & 84(10) & 120(3) & 0.057(0.001) & 0.018(0.003) & 0.000(0.002) & 0.009(0.001) & 0.090(0.001) & 0.057(0.001) \\ 
  &  &  5 & 0.3 & 128(10) & 104(2) & 0.143(0.001) & 0.035(0.003) & 0.022(0.004) & 0.031(0.001) & 0.229(0.001) & 0.167(0.001) \\ 
  &  &  6 & 1.4 & 163(10) & 92(3) & 0.047(0.001) & 0.010(0.001) & 0.000(0.003) & 0.008(0.001) & 0.071(0.001) & 0.039(0.001) \\ 
  &  &  7 & 2.5 & 151(10) & 82(4) & 0.013(0.001) & 0.008(0.001) & 0.014(0.002) & 0.003(0.001) & 0.020(0.001) & 0.010(0.001) \\ 
  &  &  8 & 3.5 & 138(11) & 106(6) & 0.012(0.001) & 0.002(0.001) & 0.010(0.002) & 0.002(0.001) & 0.013(0.001) & 0.007(0.001) \\ 
  & 2 & 2 & 5.3 & 181(11) & 44(6) & 0.005(0.001) & 0.003(0.001) & 0.004(0.002) & 0.001(0.001) & 0.006(0.001) & 0.004(0.001) \\ 
  &  &  3 & 4.2 & 184(10) & 50(5) & 0.008(0.001) & 0.000(0.001) & 0.000(0.002) & 0.001(0.001) & 0.011(0.001) & 0.006(0.001) \\ 
  &  &  5 & 2.1 & 314(10) & 83(3) & 0.032(0.001) & 0.009(0.001) & 0.005(0.002) & 0.002(0.001) & 0.052(0.001) & 0.027(0.001) \\ 
  &  &  6 & 1.1 & 300(10) & 107(2) & 0.115(0.001) & 0.020(0.002) & 0.002(0.003) & 0.022(0.001) & 0.187(0.001) & 0.131(0.001) \\ 
  &  &  7 & 0.5 & 251(10) & 98(2) & 0.068(0.001) & 0.016(0.001) & 0.000(0.003) & 0.008(0.001) & 0.101(0.001) & 0.065(0.001) \\ 
  &  &  12 & 5.4 & 277(10) & 52(5) & 0.006(0.001) & 0.000(0.001) & 0.000(0.002) & 0.002(0.001) & 0.009(0.001) & 0.002(0.001) \\ 
  &  &  13 & 6.5 & 291(11) & 50(6) & 0.004(0.001) & 0.001(0.001) & 0.000(0.002) & 6.114(0.001) & 0.007(0.001) & 0.000(0.001) \\

\end{longtable}
}
\end{center}

\end{landscape}

\section*{Appendix B}

Below, we provide examples of the quality of the spectra used in this paper. A total of four spectra are shown, two from the shock-like emission in NGC~4325, and two from the star-forming filaments in Abell~1795. These two systems cover a broad range of signal-to-noise, from very high in the centers to our detection limits at large radius. Each spectrum is divided into 5 segments in order to clearly show the quality of the fits to the various optical emission lines. Each row represents a different spatial bin along the slit. Vertical dotted lines represent the systemic velocity of the BCG.

\newcolumntype{x}[1]{>{\centering\hspace{0pt}}p{#1}}
\newcolumntype{r}[1]{>{\raggedleft\hspace{0pt}}p{#1}}

\begin{figure*}[h]
\centering
\begin{tabular}{r{2.8cm} x{2.8cm} x{2.8cm} x{2.8cm} x{2.8cm}}
\multicolumn{5}{c}{\includegraphics[width=0.9\textwidth]{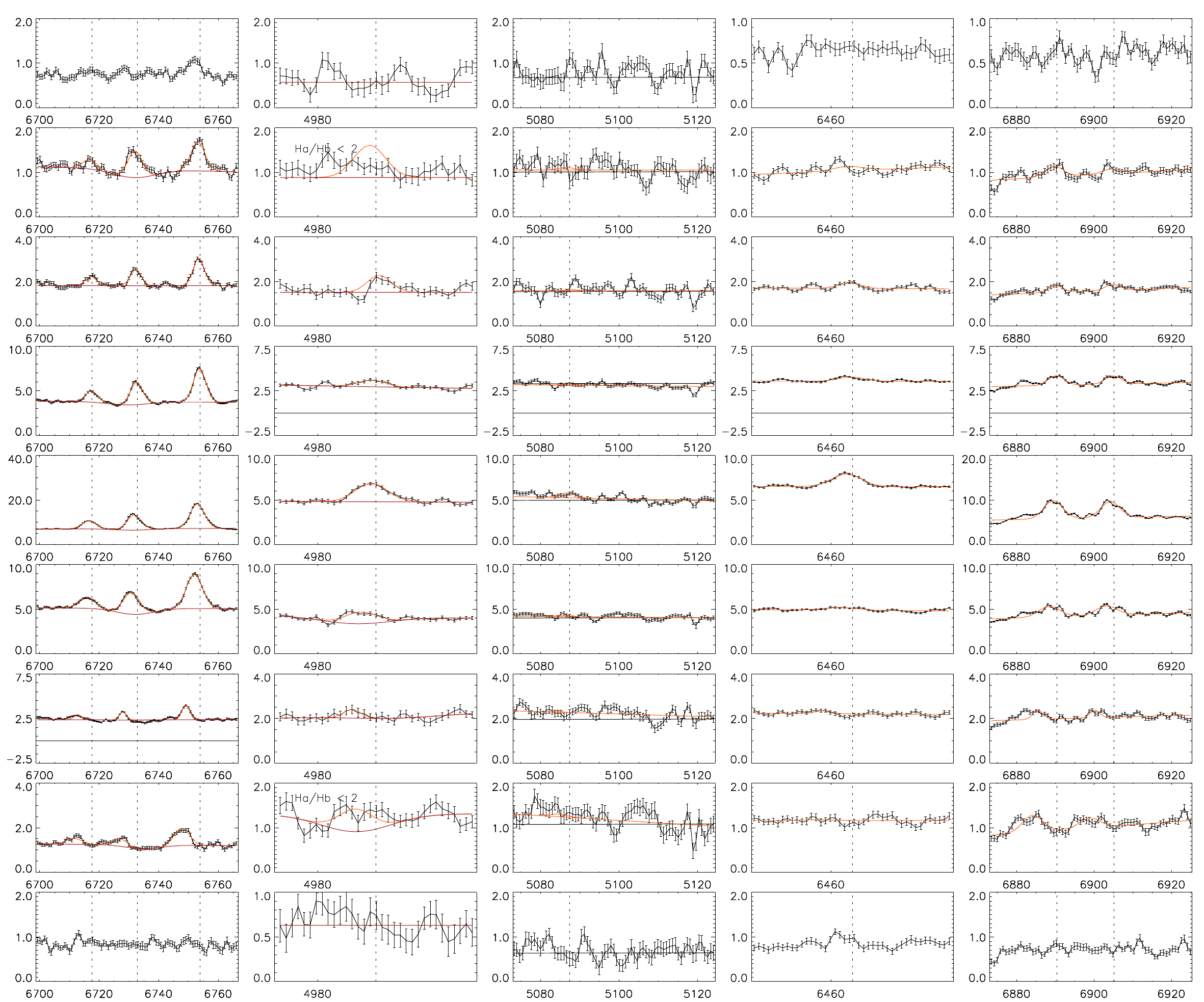}} \\
H$\alpha$ + [\ion{N}{2}] & H$\beta$ & [\ion{O}{3}] & [\ion{O}{1}] & [\ion{S}{2}]\\
\end{tabular}
\caption{NGC~4325: The broad emission lines and high [N~II]/H$\alpha$ ratios suggest that shocks may be contributing a significant amount of the total ionization in the warm gas.}
\label{}
\end{figure*}

\addtocounter{figure}{-1}
\begin{figure*}[h]
\centering
\begin{tabular}{r{3.0cm} x{3.0cm} x{3.0cm} x{3.0cm} x{3.0cm}}
\multicolumn{5}{c}{\includegraphics[width=0.999\textwidth]{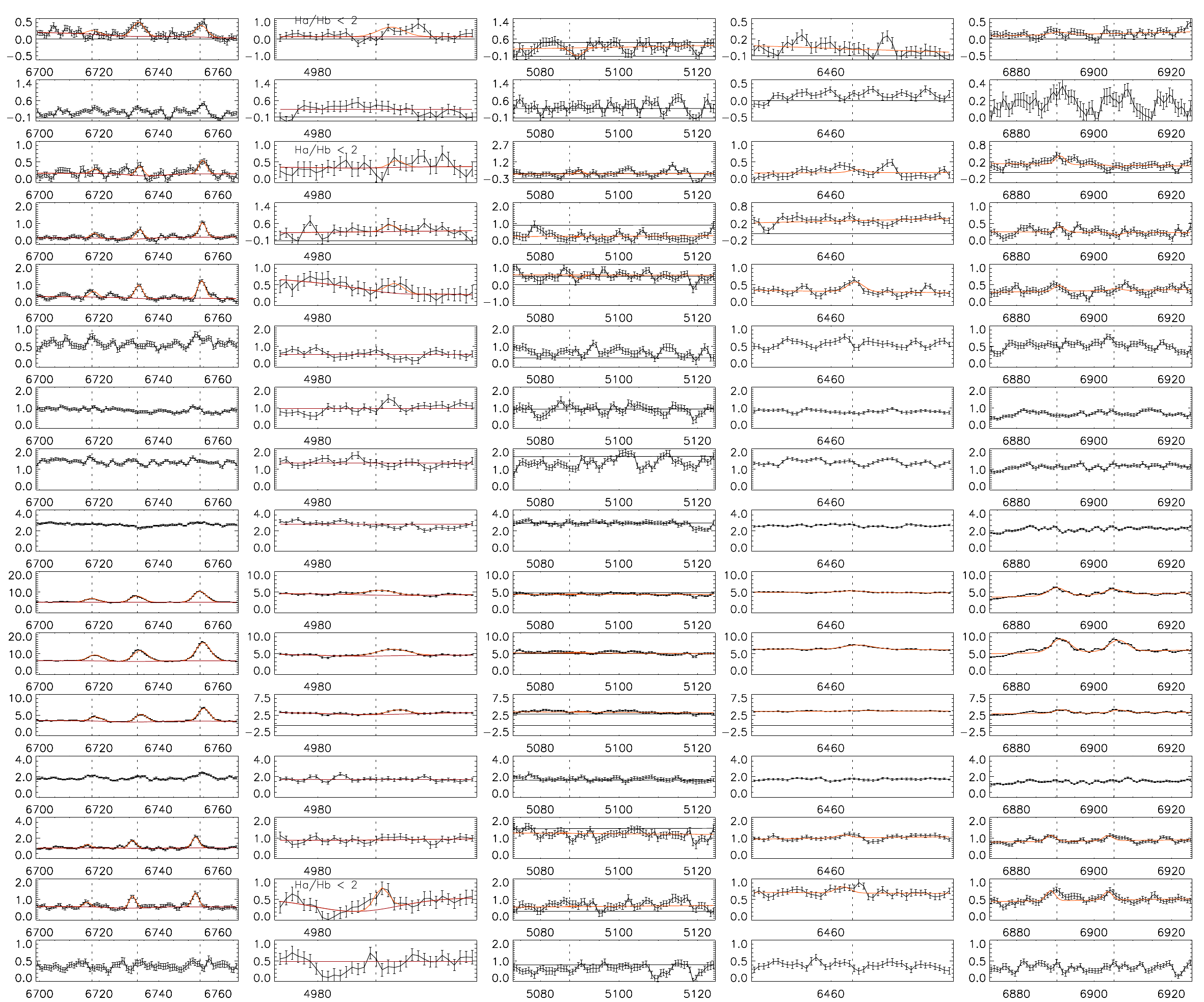}} \\
H$\alpha$ + [\ion{N}{2}] & ~~H$\beta$ & [\ion{O}{3}] & [\ion{O}{1}] & [\ion{S}{2}]\\
\end{tabular}
\caption{Continued (Slit \#2).}
\label{}
\end{figure*}

\begin{figure*}[h]
\centering
\begin{tabular}{r{3.0cm} x{3.0cm} x{3.0cm} x{3.0cm} x{3.0cm}}
\multicolumn{5}{c}{\includegraphics[width=0.999\textwidth]{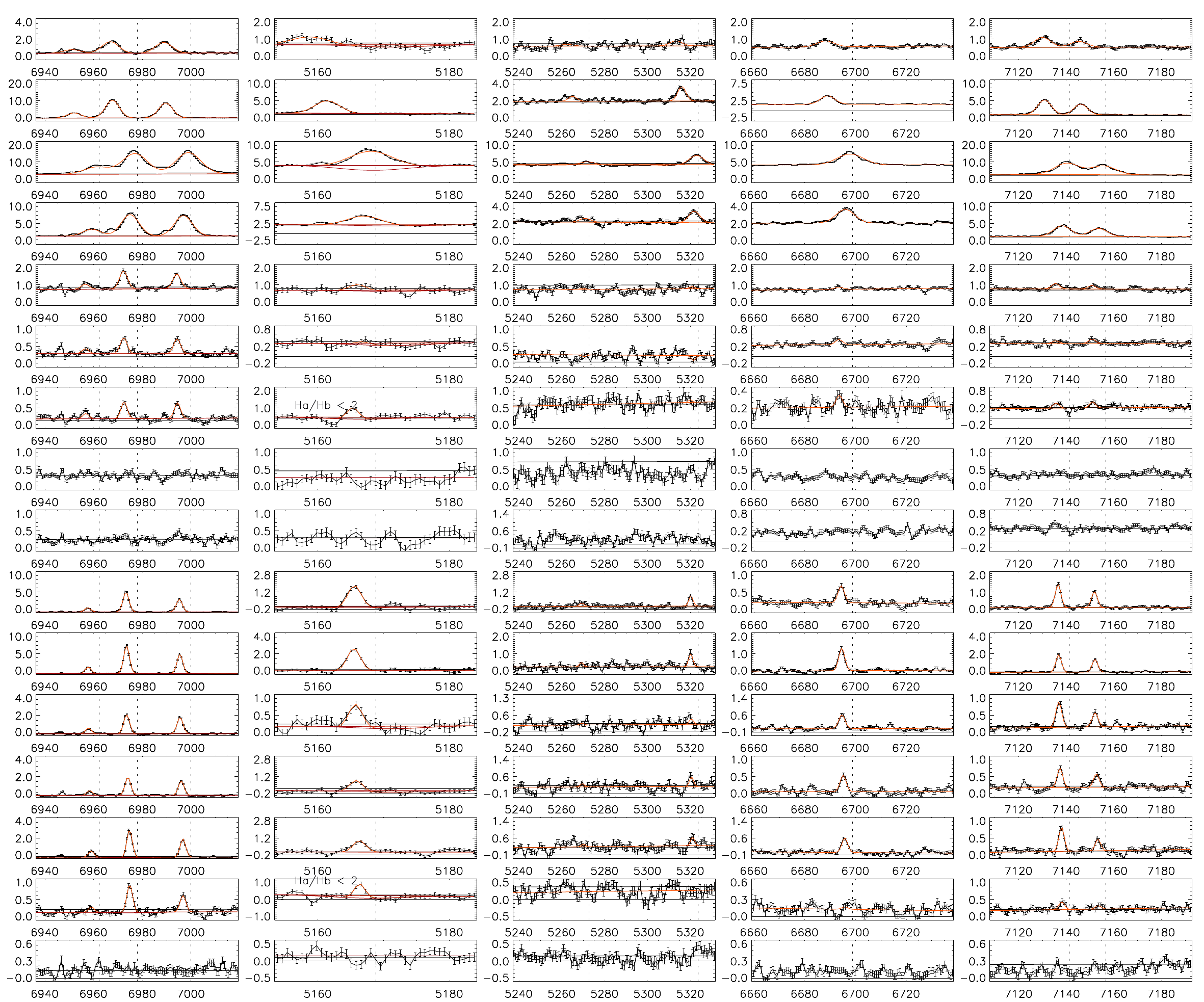}} \\
H$\alpha$ + [\ion{N}{2}] & ~~H$\beta$ & [\ion{O}{3}] & [\ion{O}{1}] & [\ion{S}{2}]\\
\end{tabular}
\caption{Abell~1795: The narrow emission lines and small [N~II]/H$\alpha$ suggests that emission in these filaments is mostly due to star formation. The high [O~I] luminosity requires a small fractional contribution ($\sim$20\%) from shocks.}
\label{}
\end{figure*}

\addtocounter{figure}{-1}
\begin{figure*}[h]
\centering
\begin{tabular}{r{3.0cm} x{3.0cm} x{3.0cm} x{3.0cm} x{3.0cm}}
\multicolumn{5}{c}{\includegraphics[width=0.999\textwidth]{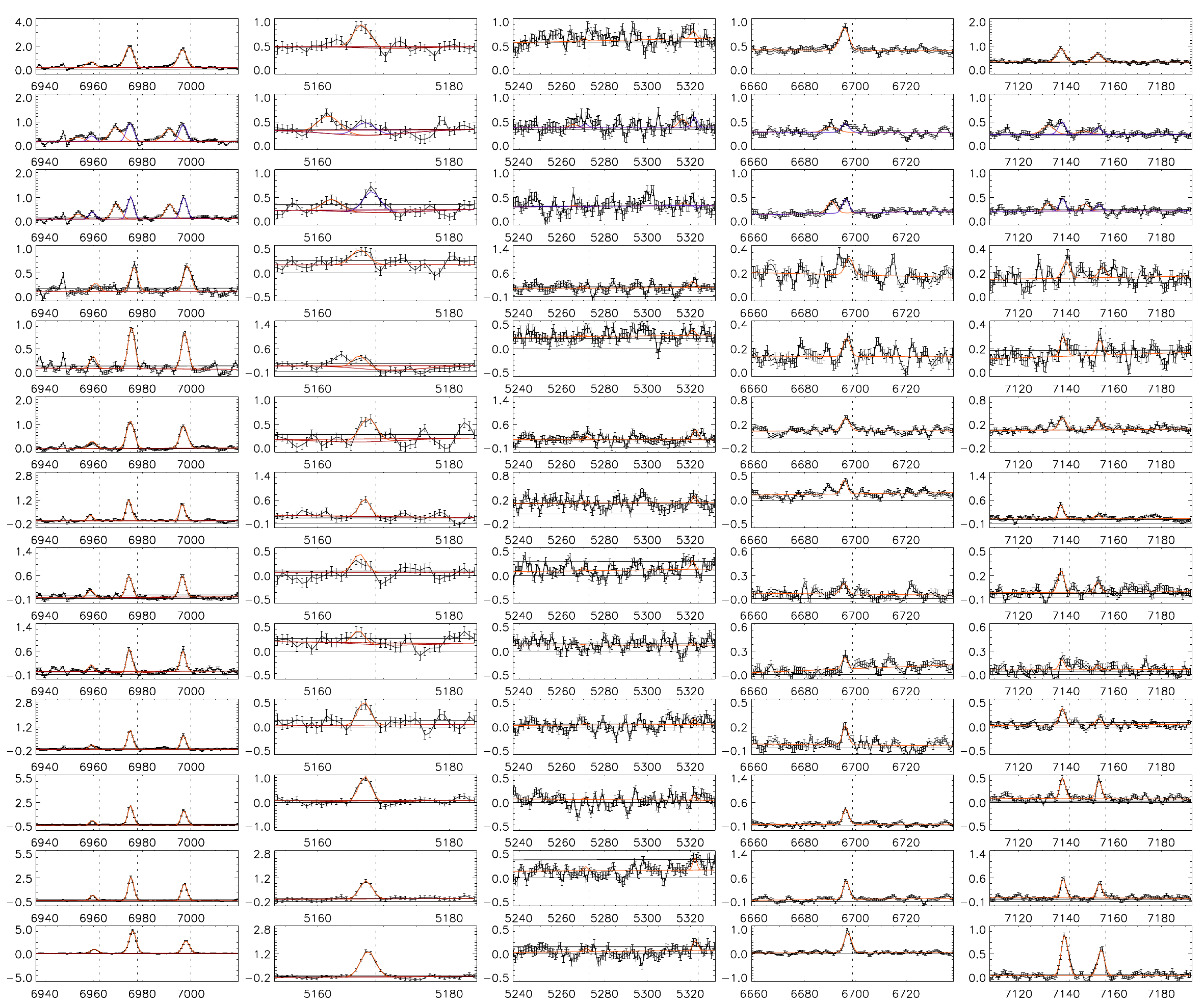}} \\
H$\alpha$ + [\ion{N}{2}] & ~~H$\beta$ & [\ion{O}{3}] & [\ion{O}{1}] & [\ion{S}{2}]\\
\end{tabular}
\caption{Continued (Slit \#2).}
\label{}
\end{figure*}

\end{document}